\documentclass[showpacs,prl,onecolumn,aps,superscriptaddress,preprintnumbers,nofootinbib]{revtex4}
\usepackage[T1]{fontenc}
\usepackage[latin9]{inputenc}
\usepackage{amsmath,amssymb}
\usepackage{epsfig}
\usepackage{bm} 
\usepackage{graphicx}
\usepackage{mathrsfs}
\usepackage{amsmath}
\usepackage{amsfonts}
\usepackage{epstopdf}
\usepackage{color}
\usepackage{pifont}
\usepackage{relsize}
\usepackage{mathtools}
\def\slashchar#1{\setbox0=\hbox{$#1$}     		% set a box for #1
   \dimen0=\wd0                                 	% and get its size
   \setbox1=\hbox{/} \dimen1=\wd1               	% get size of /
   \ifdim\dimen0>\dimen1                        	% #1 is bigger
      \rlap{\hbox to \dimen0{\hfil/\hfil}}      	% so center / in box
      #1                                        	% and print #1
   \else                                        	% / is bigger
      \rlap{\hbox to \dimen1{\hfil$#1$\hfil}}   	% so center #1
      /                                         	% and print /
   \fi}

\renewcommand{\vec}{\boldsymbol}
\newcommand{\beq}{\begin{equation}}
\newcommand{\eeq}{\end{equation}}
\newcommand{\bea}{\begin{eqnarray}}
\newcommand{\eea}{\end{eqnarray}}
\newcommand{\ba}{\begin{array}}
\newcommand{\ea}{\end{array}}

\def\eq#1{{Eq.~(\ref{#1})}}
\def\fig#1{{Fig.~\ref{#1}}}
\newcommand{\bas}{\bar{\alpha}_S}

\newcommand{\nn}{\nonumber}
\newcommand{\bg}{ \bar{\gamma}}

\newcommand{\Lb}{\left(}
\newcommand{\Rb}{\right)}
\newcommand{\h}{\frac{1}{2}}
\newcommand{\rv}{\vec{r}}

\newcommand{\bv}{\vec{b}}
\newcommand{\pom}{I\!\!P}

\newcommand{\intl}{\int\limits}

\begin{document}

%%%%%%%%%%%%%%%%%%%%%%%%%%%%%%%%%%%%%%%%%%%%%%%%%%%%%%%%%%%%%%%%%%%%%%%
\title{ Dipole-dipole scattering: summing large Pomeron loops\\ in non-linear evolution with leading twist kernel }
\author{Eugene Levin}
\email{leving@tauex.tau.ac.il}
\affiliation{Department of Particle Physics, Tel Aviv University, Tel Aviv 69978, Israel}

\date{\today}


\begin{abstract}
 It is shown in this paper that  the QCD equations for dipole density have the natural solution: the 'fan' diagrams of the Pomeron calculus. We found the dipole densities 
 comparing 
 the analytic solution to the 
Balitsky-Kovchegov (BK) equation for the simplified leading twist kernel with    the $t$ channel unitarity.
Using these densities we  calculate the contributions of large Pomeron loops to dipole-dipole scattering at high energies.  Applying the Abramovsky,Gribov and Kancheli cutting rules we found  that  the produced gluons are distributed accordingly the KNO (Koba, Nielsen and Olesen) law which leads to the entropy $S_E = \ln(x G(x,Q^2))$ in an agreement with Kharzeev - Levin predictions.

 \end{abstract}
\maketitle

\vspace{-0.5cm}
\tableofcontents

%%%%%%%%%%%%%%%%%%%%%%%%%%%%%%%%%%%%%%%%%%%%%%%%%%%%
\section{Introduction}
 %%%%%%%%%%%%%%%%%%%%%%%%%%%%%%%%%%%%%%%%%%%%%%%%%%%% 
 In this paper we continue \cite{LE1,LEDIDI,LEDIA,LEAA}  our attempts to   sum  large
  BFKL Pomeron loops\cite{BFKL}\footnote{BFKL stands for Balitsky, Fadin,Kuraev and Lipatov.} for dipole-dipole scattering at high energies. The main goal of this paper is to expand our methods, suggested in our previous publications\cite{LEDIDI,LEDIA,LEAA}, to a wider range of energy going from ultra high energies  to high energies in the saturation region and   to check the procedure of summing the asymptotic series for the Pomeron exchanges for the scattering amplitude. It is well known that the BFKL  parton cascade leads to the non-linear Balitsky-Kovchegov (BK) equation \cite{B,K}. In Ref.\cite{LETU} the analytical solution to this equation in the saturation region was found:
 \beq\label{I1}
 N^{\rm DIS} \Lb z= r^2 Q^2_s\Lb Y, b\Rb\Rb\,\,=\,\,1\,\,-\,\,C(z)\exp\Lb - \frac{z^2}{2\,\kappa}\Rb
 \eeq 
 where $z = \ln\Lb r^2 Q^2_s(Y)\Rb$\footnote{We will discuss the definition of $z$  and of $\kappa$ in more detail  below in \eq{zz} , \eq{ZA}  and \eq{GACR}.}.   
 The main idea of this paper is the same as   our previous ones \cite{LEDIDI,LEDIA,LEAA}. We wish  to reconcile this solution with the fact that the scattering amplitude can be presented as the asymptotic series of the multi Pomeron exchanges. In doing so we will find the parton densities in QCD and will be able to calculate the sum of large Pomeron loops in  dipole-dipole scattering.
 
 In our previous works we found that the sum of the large Pomeron loops leads to factor 
 $\exp\Lb - \frac{z^2}{4\,\kappa}\Rb $ in the scattering amplitudes of dipole-dipoles, dipole-nucleus and nucleus-nucleus. In this paper we are going to discuss the smooth function $C\Lb z\Rb$ in \eq{I1} for dipole-dipole scattering.   At first sight it  looks that we are after  a minor point.  However,  we wish to address the principle problem in this paper: the method of summing of asymptotic Pomeron series. In our previous paper we assumed the specific  approach to this resummation and in this paper we expect to check this method  with the approach which is based on the specific properties of BK equation.

  However at the moment we can do this for the simplified BFKL kernel: the BFKL kernel in the leading twist. 
 
   Summing Pomeron loops  has been one of the  difficult problems in the Color Glass Condensate (CGC) approach, without solving which  we cannot consider the dilute-dilute and dense-dense parton densities collisions. However, in spite of intensive work 
\cite{BFKL,KOLEB,KLLN,MUSA,LETU,LELU,LIP,KO1,LE11,RS,KLremark2,SHXI,KOLEV,nestor,LEPRI,LMM,LEM,IAMU,IAMU1,MUT,IIML,LIREV,LIFT,GLR,GLR1,MUQI,MUDI,Salam,NAPE,BART,BKP,MV, KOLE,BRN,BRAUN,B,K,KOLU,JIMWLK1,JIMWLK2,JIMWLK3, JIMWLK4,JIMWLK5,JIMWLK6,JIMWLK7,JIMWLK8,AKLL,KOLU11,KOLUD,BA05,SMITH,KLW,KLLL1,KLLL2,kl,LEPR,LE1,LEDIDI,LEDIA,LEAA}, this problem  has  not been solved.  

\begin{figure}[ht]
    \centering
  \leavevmode
      \includegraphics[width=7cm]{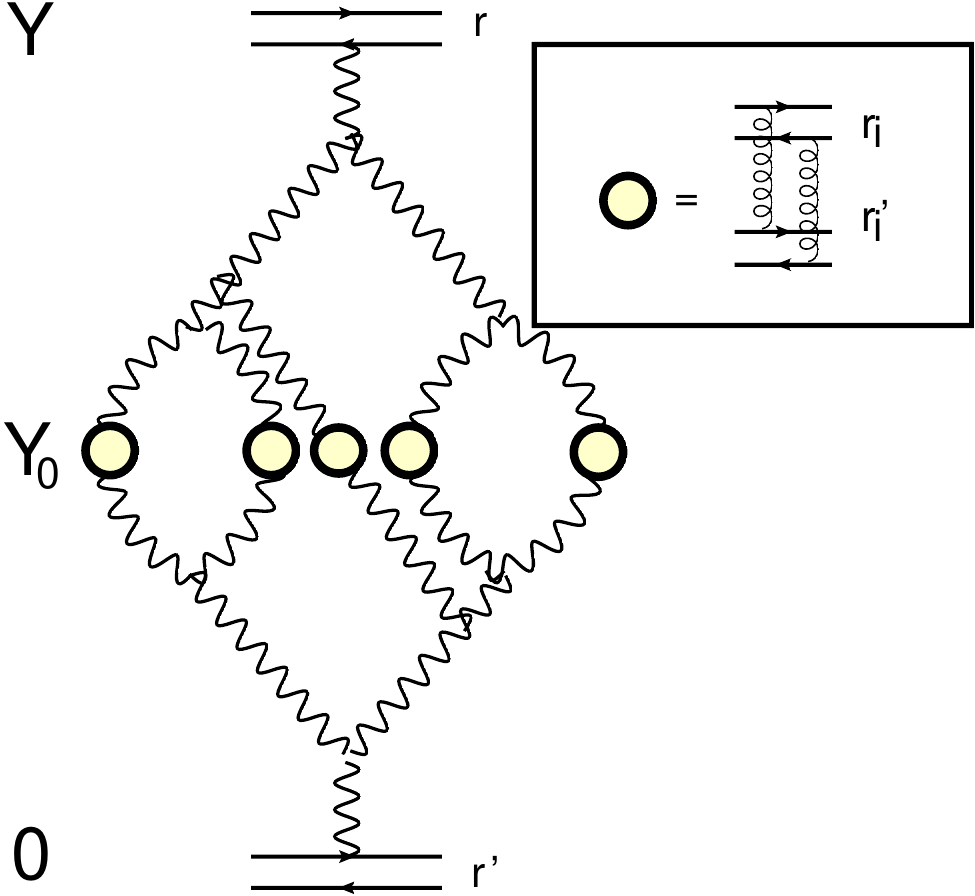}  
      \caption{ Summing  large Pomeron loops. The wavy lines denote the  BFKL Pomeron exchanges.  The circles denote the amplitude $\gamma$.  }
\label{mpsi}
   \end{figure}

 We sum the large Pomeron loops using the $t$-channel unitarity, which has been rewritten in the convenient form for the dipole approach to CGC in Refs.\cite{MUSA,Salam,IAMU,IAMU1,KOLEB,MUDI,LELU,KO1,LE11,AKLL}(see \fig{mpsi}).
 $ N\Lb Y, r,r' ;  \vec{b}\Rb$ is scattering amplitude for the interaction  
 of dipole with size $r$, impact parameter $b$ and rapidity $Y$ with the dipole of the size $r'$, which is at the rest. The analytic expression for the imaginary part of this amplitude
takes the form
       \cite{LELU,KO1,LE1}:  \bea \label{MPSI}
     && N\Lb Y, r, r' ;  \vec{b}\Rb\,=\\
     &&\,\sum^\infty_{n=1}\!\!\Lb -1\Rb^{n+1}\!n!\!\!\intl \!\! \prod^n_i d^2 r_i\,d^2\,r'_i\,d^2 b'_i \!\!
     \intl \!\!d^2 \delta b_i\, \gamma^{BA}\Lb r_1,r'_i, \vec{b}_i -  \vec{b'_i}\equiv \delta \vec{b} _i\Rb 
    \,\,\rho_n\Lb Y - Y_0, \vec{r}, \vec{b}, \{ \vec{r}_i,\vec{b}_i\}\Rb\,\rho_n\Lb Y_0,\vec{r}', \vec{b}'=0, \{ \vec{r}'_i,\vec{b}'_i\}\Rb \nn
      \eea
  $\gamma^{BA}$ is the scattering amplitude of two dipoles in the Born approximation of perturbative QCD.  The dipole densities $\rho_i Y , \{ \vec{r}_i,\vec{b}_i\}$ have been introduced in Ref.\cite{LELU}  as follows:
\beq \label{PD}
\rho_n(r, b, r_1, b_1\,
\ldots\,,r_n, b_n; Y\,-\,Y_0)\,=\,\frac{1}{n!}\,\prod^n_{i =1}
\,\frac{\delta}{\delta
u_i } \,Z\left(Y\,-\,Y_0;\,[u] \right)|_{u=1}
\eeq
  where  the generating functional $Z$ is
  \beq \label{Z}
Z\Lb Y, \vec{r},\vec{b}; [u_i]\Rb\,\,=\,\,\sum^{\infty}_{n=1}\int P_n\Lb Y,\vec{r},\vec{b};\{\vec{r}_i\,\vec{b}_i\}\Rb \prod^{n}_{i=1} u\Lb \vec{r}_i\,\vec{b}_i\Rb\,d^2 r_i\,d^2 b_i
\eeq
 where $u\Lb \vec{r}_i\,\vec{b}_i\Rb \equiv\,u_i$ is an arbitrary function and $P_n$ is the probability to have $n$ dipoles with the  given kinematics.
 The initial and  boundary conditions for the BFKL cascade which stems from one dipole has 
the following form for the functional $Z$:
 
 \begin{subequations}
\bea
Z\Lb Y=0, \vec{r},\vec{b}; [u_i]\Rb &\,\,=\,\,&u\Lb \vec{r},\vec{b}\Rb;~~~~~~~~Z\Lb Y, r,[u_i=1]\Rb = 1;\label{ZIC}\\
\rho_1\Lb Y=0,   r,b, r_1,b_1\Rb\,\,&=&\,\,\delta^{(2)}\Lb \vec{r} - \vec{r}_1\Rb \delta^{(2)}\Lb \vec{b} - \vec{b}_1\Rb ;~~~~~\rho_n\Lb 
 Y=0, \vec{r},\vec{b}; [r_i, b_i]\Rb \,=\,0 ~ \mbox{at}~\,n\geq 2;\label{ZSR}
\eea
\end{subequations}
  In \eq{MPSI} $\vec{b}_i\,\,=\,\,\vec{b} \,-\,\vec{b'}_i$.

            %%%%%%%%%%%%%%%%%%%%%%%%%%%%%%%%%%%%%%%%%
%%%%%%%%%%%%%%%%%%%%%%%%%%%%%%%%%%%%%%%%%%    
  As one can see from \eq{I1} the 
 BK equation  leads to a new dimensional scale: saturation momentum\cite{GLR}  which has the following $Y$ dependence\cite{GLR,MUT,MUPE} \footnote{  Actually \cite{MUT,MUPE}  $Q^2_s\Lb Y, b\Rb\,\,=\,\,Q^2_s\Lb Y=0, b\Rb \,e^{\bas\,\kappa \,Y\,-\,\,\frac{3}{2\,\gamma_{cr}} \ln Y }$ but through the paper we neglect the contribution  $\propto \ln Y$ in the saturation momentum for the sake of simplicity of the presentation. This simplificatio does not change any of our results and could be easily taken off.}
:
 \beq \label{QS}
 Q^2_s\Lb Y, b\Rb\,\,=\,\,Q^2_s\Lb Y=0, b\Rb \,e^{\bas\,\kappa \,Y }
 \eeq 
 where $Y=0$ is the initial value of rapidity and $\kappa$ and $\gamma_{cr}$   are determined by the following equations\footnote{$\chi\Lb \gamma\Rb$ is the BFKL kernel\cite{BFKL} in anomalous dimension ($\gamma$) representation.$\psi$ is the Euler psi -function (see Ref.\cite{RY} formula {\bf 8.36}). }:
  \beq \label{GACR}
\kappa \,\,\equiv\,\, \frac{\chi\Lb \gamma_{cr}\Rb}{1 - \gamma_{cr}}\,\,=\,\, - \frac{d \chi\Lb \gamma_{cr}\Rb}{d \gamma_{cr}}~
\eeq
where $\chi\Lb \gamma\Rb$ is given by
\beq \label{CHI}
\omega\Lb \bas, \gamma\Rb\,\,=\,\,\bas\,\chi\Lb \gamma \Rb\,\,\,=\,\,\,\bas \Lb 2 \psi\Lb 1\Rb \,-\,\psi\Lb \gamma\Rb\,-\,\psi\Lb 1 - \gamma\Rb\Rb\eeq 

 In \eq{MPSI} $\rho_n$ are functions  of the rapidities and the sizes of dipoles $\vec{r}_i = \vec{x}_i - \vec{y}_i$ and their impact parameters $\vec{b}_i = \h\Lb \vec{x}_i + \vec{y}_i\Rb$, where 
$\vec{x}_i$ and $\vec{y}_i$ are the coordinates of the quark and antiquark of the dipole $i$\footnote{In the paper we also use  different notations for $\vec{x}_i $ and $ \vec{y}_i$: $ \vec{x}_i $ and $\vec{x}_k$.}. In \eq{MPSI}  we put $r, b$  among the arguments of $\rho_n$, however, below we will often use $\rho_n\Lb\{ \vec{r}_i,\vec{b}_i\}\Rb$ since the dependence on $r$ and $b$ comes only from the initial condition of \eq{ZSR}.

In this paper we consider the scattering amplitudes and $\rho_n$ in the saturation region. In this region these observables show the geometric scaling behaviour \cite{GS} when  the number of variables  are reduced.  It turns out that they depend on 

 \begin{subequations}
 \bea 
N\Lb Y, r, r' ;  \vec{b}\Rb\,=\,N\Lb z\Rb & & z\,\,=\,\,\bas\, \kappa \,Y \,\,+\,\,\xi_{r,r'}\,; \label{zz}\\
\rho_n\Lb Y - Y_0, \vec{r}, \vec{b}, \{ \vec{r}_i,\vec{b}_i\}\Rb=\rho\Lb\{z_i\}\Rb & & 
z_i\,=\,\bas\,\kappa \,\Lb Y \,-Y_0\Rb \,\,+\,\,\xi_{r,r_i}\,; \label{zi}\\
\rho_n\Lb Y_0,\vec{r}', \vec{b}'=0, \{ \vec{r}'_i,\vec{b}'_i\}\Rb= \rho_n\Lb \{ z'_i\}\Rb & & 
z'_i\,=\,\bas \,\kappa\,Y_0 \,\,+\,\,\xi_{r'_i,r'}\,; \label{zip}
\eea
\end{subequations}
The variable $\xi$ in these equations will be defined below in \eq{XI} from the solution to the BFKL equation in perturbative QCD. For \eq{zi} $\xi_{r, r_i}$ depends on the 
sizes of two dipoles at rapidity $Y$ and $Y_0$ \footnote{ For \eq{zz} the rapidities are equal to $Y$ and 0. For \eq{zip} they are $Y_0$ and 0.},
which are shown as the indices of $\xi$.
The difference between the  impact parameters of these dipoles are not  in our notations. If this difference is  smaller than the largest size  $\xi_{r,r_i} = \ln \Lb r^2/r^2_i\Rb$ as it has been shown in \eq{I1}.

\begin{figure}[ht]
    \centering
  \leavevmode
      \includegraphics[width=7cm]{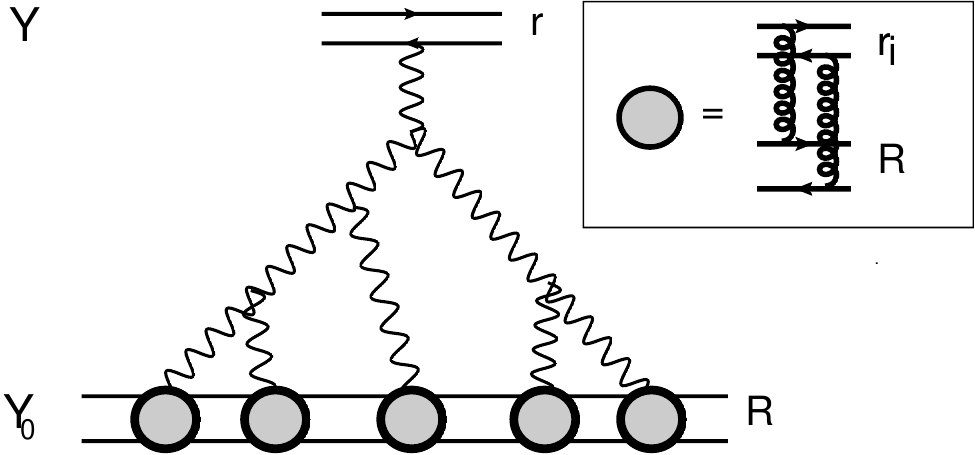}  
      \caption{The non-linear BK equation. The wavy lines denote the  BFKL Pomeron exchanges.  The circles  denote the amplitude $\gamma$ for interaction of the dipole with the target.  }
\label{mpsi1}
   \end{figure}

For the interaction with nuclei (see \fig{mpsi1}) the scaling variable has the following form
\beq \label{ZA}
z_A\,\,=\,\,\bas \,\kappa\,\,Y \,\,+\,\,\xi_{r,R_A}
\eeq
$R$ is the size of the typical dipole in a nucleus. Since for the scattering with nucleus the impact parameters are large $\xi_{r,R_A}$ is equal to $\xi_{r,R_A} =\ln\Lb \frac{r^2\,R^2_A}{b^4}\Rb$ from \eq{XI}. However, for the interaction with nuclei we need to take into account the non-perturbative dependence of the saturation scale on $b$, which we will discuss in section IV. In \eq{I1} we consider the BK amplitude for the interaction with nucleus and $Q_s $ in the definition of $z$ is equal $Q_s\Lb Y \Rb = \frac{1}{R_A} \exp\Lb \h \bas \,\kappa\, \,Y\Rb$.
           %%%%%%%%%%%%%%%%%%%%%%%%%%%%%%%%%%%%%%%%%
%%%%%%%%%%%%%%%%%%%%%%%%%%%%%%%%%%%%%%%%%%   

In the next section we review the Balitsky-Kovchegov (BK) non-linear equation for the simplified, leading twist BFKL kernel. We will discuss the solution to this equation which  has  a simple analytic form for this  kernel. In section III we show how this solution can be reconciled with the BFKL Pomeron calculus. We demonstrate that the natural  solution to the QCD evolution equations for dipole densities $\rho_n$ is the 'fan' BFKL Pomeron diagrams. In this section we also discuss how $\rho_n$   contribute  to summing  large Pomeron loops.

In section IV
 we find the BK  scattering amplitude as a sum of many Pomerons exchanges and in doing so we reconstruct the  densities $\rho_n$. We also discuss and find the method  to sum the asymptotic series of the multi Pomeron exchanges that lead to the scattering amplitude.  It should be noted right now that the method  of summing the  asymptotic Pomeron series turns out to be quite different from the one in Ref.\cite{LEDIDI}. The attractive feature of the found Pomeron series is its similarity to the series in the one dimensional model which satisfy both $t$ and $s$ unitarity\cite{MUSA,KLLN}   and which is solved exactly.
 
  In section V we use the advantage of the multi  Pomeron series for the scattering amplitude and calculated the multiplicity distributions based on s-channel unitarity for the BFKL Pomeron and the AGK cutting rules.  It is instructive to note that in spite of the same entropy as in Ref.\cite{LEDIDI} the multiplicity distribution turns out to be quite different. This fact is a reflection of a different way of summation of the asymptotic Pomeron series.
 
In section VI we calculate the scattering amplitude for the dipole-dipole scattering using \eq{MPSI} and in section VII we present  our calculations  of the multiplicity distribution for the dipole-dipole amplitude. In section VIII we show that the entropy of the produced gluons for both processes give   $S_E \,=\,\ln(xG(x,Q^2))$ where $xG$ is the gluon structure function. Therefore it confirms the result of Ref.\cite{KHLE}. However,  the multiplicity distribution turns out to be different.

In conclusion we summarize our results and discuss possible flaw in our approach,

         %%%%%%%%%%%%%%%%%%%%%%%%%%%%%%%%%%%%%%%%%%%%%%%%%%%%
    \begin{boldmath}
    \section{Balitsky-Kovchegov non-linear equation for the leading twist BFKL kernel } 
    \end{boldmath}
    
%%%%%%%%%%%%%%%%%%%%%%%%%%%%%%%%%%%%%%%%%%%%%%%%%%%%  
 The scattering     matrix (S-matrix) of the colourless dipole with the size $x_{01}$ satisfies the Balitsky-Kovchegov (BK) non-linear equation\cite{B,K}:
 \beq \label{BK}
 \frac{\partial S_{01}}{\partial Y}\,=\,\bas\int \frac{d^2\,x_{02}}{2 \pi} \frac{ x^2_{01}}{x^2_{02}\,x^2_{12}}\Big\{ S_{02}\,S_{12} - S_{01}\Big\}
 \eeq
 where $S_{ik}=S\Lb Y, \vec{x}_{ik},\vec{b}\Rb$\footnote{For simplicity we assume that $b\gg x_{i,k} $. The  equation without this assumption  has a bit more cumbersome form and can be found in Ref.\cite{KOLEB}, for example.}   is the scattering matrix of the dipoles with size $x_{ik}$ and with rapidity $Y$ at the impact parameter $\vec{b}$.
 For the full BFKL kernel the solution at ultra high energy\cite{LETU}  comes from \eq{BK} assuming that $S_{02}(S_{12})$ are small in the saturation region. In this region  \eq{BK} degenerates to
  \beq \label{BKA}
 \frac{\partial S_{01}}{\partial Y}\,=\,-\bas\intl^{x_{01}}_{\frac{1}{Q_s}} \frac{d^2\,x_{02}}{2 \pi} \frac{ x^2_{01}}{x^2_{02}\,x^2_{12}} S_{01}
 \eeq

For the geometric scaling solution inside the saturation region\cite{GS} \eq{BKA} can be rewritten as follows\cite{LETU}:
\beq \label{BKA2}
\kappa \frac{d S_{01}\Lb z\Rb}{d\, z}=  - z \,\,S_{01}\Lb z\Rb
\eeq
with the solution:
\beq \label{BKA3}
S_{01}\Lb z\Rb= C(z) \exp\Lb - \frac{z^2}{2\,\kappa}\Rb
\eeq
where  $C(z)$ is a smooth function of $z$. $z$ is given by \eq{zz}. In this derivation we  consider that the reggeization of  gluon gives $ \bas \ln\Lb x^2_{01} Q^2_s(Y)\Rb = \bas\,z$. Unfortunately we are not ready at the moment to provide solution for $C(z)$ in \eq{BKA3} for the general BFKL kernel. We will specify the solution in the entire region of $z$ in the saturation region for 
  the simplified BFKL kernel    suggested in     Ref.\cite{LETU} . This kernel describes the high energy asymptotic solution of the nonlinear BK equation and leads to the geometric scaling behaviour.: 
    \bea \label{SIMKER}
\chi\Lb \gamma\Rb\,\,=\,\, \left\{\begin{array}{l}\,\,\,\frac{1} {1\,-\,\gamma}\,\,\,\,\,\,\,\,\,\,\mbox{for}\,\,z = \xi_{r,R} + \kappa\,\bas\,Y \,>\,0,\,\,\,\,\,\,\mbox{summing} \Lb z\Rb^n;\\ \\
\,\,\,\frac{1}{\gamma}\,\,\,\,\,~~~~~~\mbox{for}\,\,\,z \,<\,0,\,\,\,\,\,~~~~~~\,\,\,\,\,~~~~~~~~~~~\mbox{summing}
\Lb \xi\Rb^n;\\  \end{array}
\right.
\eea
 where  $\kappa = 4 $ for this kernel and
 $\xi_{r,R}$ will be defined in \eq{XI} below.

 Since this kernel sums log contributions it corresponds to leading twist   term of the full BFKL kernel. It has a very simple form in the coordinate representation \cite{LETU}:
 
\beq \label{K2}
\intl \, \displaystyle{ K\Lb \vec{r}',\vec{r} - \vec{r'}|\vec{r}\Rb}\,d^2 r' \,\rightarrow
\,\frac{\bas}{2}\!\!\!\!\! \!\!\!\!\! \intl^{r^2}_{1/Q^2_s(Y,b)}\!\!\!\!\!\!  \frac{ d r'^2}{r'^2}\,\,+\,\,
\frac{\bas}{2}\!\!\!\!\! \!\!\!\!\! \intl^{r^2}_{1/Q^2_s(Y, b)}\!\!\!\!\!\!\! \frac{ d |\vec{r} - \vec{r}'|}{|\vec{r}  - \vec{r}'|^2}\,\,
 =\,\,\frac{\bas}{2}\, \intl^{\xi}_{\xi_s} d \xi_{r'} \,\,+\,\,\frac{\bas}{2}\, \intl^{\xi}_{\xi_s} d \xi_{\vec{r} - \vec{r}'}\eeq
 where $\xi_{r'} \,=\,\ln \Lb \frac{r'^2\,r^2_1}{b^4}\Rb$ for $b \,>\,r,r_1$
 and $\xi_s\,=\,\kappa\, \bas\,Y$ for the scattering of the  dipole $r'$ with the dipole $r_1$. The general form of $\xi(r')$ we will discuss below in \eq{XI}. Recall that we consider the scattering matrix for the dipoles of size $r'$ which scatters with the dipole of size $r_1$ at impact parameter $b$.
 Note, that  the  logarithms
   originate from the decay of a large size dipole, into one small
 size dipole  and one large size dipole\cite{LETU}.  However, the size of the
 small dipole is still larger than $1/Q^2_s  = \frac{b^4}{r^2_1}\,e^{-\kappa Y}  $.      Using \eq{K2} and \eq{BK} we can obtain the non-linear equation for the scattering amplitude $N_{i,k} = 1 - S_{ik}= 1 - S\Lb Y, \vec{x}_{ik},\vec{b}\Rb$ in the simple form:
 
 \beq \label{BKS}
 \frac{ \partial\, N\Lb Y, \xi,b\Rb }{\partial \,Y}\,\,=\,\,\bas \Lb 1 - N\Lb Y,\xi, b\Rb \Rb\int^\xi_{\xi_s} d \xi' N\Lb Y,\xi', b\Rb
\eeq
and for the geometric scaling solution:
 \beq \label{BKS1}
 4\frac{ d \,N\Lb z\Rb}{d \,z}\,\,=\,\, \Lb 1 - N(z) \Rb\int^z_0 d z' N\Lb z'\Rb
\eeq

This equation gives the BK equation for our kernel. The linear BFKL equation is: 
\beq \label{BKS10}
 4\frac{ d \,N\Lb z\Rb}{d \,z}\,\,=\,\,\int^z_0 d z' N\Lb z'\Rb
\eeq
The general solution for this equation has the following form:
 \beq \label{BKS11}
N_{\rm lin}\Lb z \Rb\,\,=\,\,C_1 e^{\h \,z}\,\,+\,\,C_2 e^{ - \h\,z}
\eeq
The second solution decreases at large $z$ and could be important only at the border of the saturation region. The first solution is the BFKL Pomeron exchange which satisfy the boundary condition at $z=0$\cite{GLR,MUT}, viz. 
\beq \label{BKS2}
G_{\pom}\Lb z \Rb  = N_0 e^{\bg \,z}\eeq
where $\bg = 1 - \gamma_{cr}=\h$ , $N_0$ is a constant.  $G_{\pom} \Lb z \Rb $ is the Green's function of the BFKL Pomeron exchange. For  exchanges of $n$ BFKL Pomeron we have the Green's function in the form:
\beq \label{BKS3}
G_{n \pom} = \Lb G_{\pom}\Lb z \Rb \Rb^n\eeq
 $z$ in \eq{BKS1} and \eq{BKS3}  is determined  by \eq{zz} (see \fig{mpsi1})
where
$\xi_{r, r'}$ we can find from  the eigenfunction of the BFKL equation with the eigenvalue $\bas \chi\Lb \gamma\Rb$
 (the scattering amplitude of two dipoles with sizes $r$ and $r'$) which  has the
 following form \cite{LIP}
\beq \label{XI}
\phi_\gamma\Lb \vec{r} , \vec{r}', \vec{b}\Rb\,\,\,=\,\,\,\Lb \frac{
 r^2\,r'^2}{\Lb \vec{b}  + \h(\vec{r} - \vec{r}')\Rb^2\,\Lb \vec{b} 
 -  \h(\vec{r} - \vec{r}')\Rb^2}\Rb^\gamma =e^{\gamma\,\xi_{r, r'} }~~\mbox{with}\,\,0 \,<\,Re\,\gamma\,<\,1
 \eeq 

\eq{BKS} can be rewritten  in a simpler form for $\Omega\Lb \xi_s,\xi\Rb $: $ N\Lb Y, \xi, b\Rb= 1\,\,-\,\,\exp\Lb - \Omega\Lb \xi_s,\xi\Rb\Rb$, with $\xi_s = \bas\,\kappa\,Y$. It takes 
the form:
\beq \label{BKS4}
\frac{\partial^2\, \Omega\Lb \xi_s,\xi\Rb}{\partial \,\xi_s\,\,\partial\,\xi}\,\,=\,\,\frac{1}{4}\Lb 1\,\,-\,\,e^{ -\Omega\Lb \xi_s,\xi\Rb}\Rb
\eeq

\eq{BKS4} has the geometric scaling solution (see formula {\bf 3.5.3} in Ref.\cite{MATH}), which has the following implicit form:
\beq \label{BKS5}
\sqrt{2}\intl^{\Omega\Lb z \Rb}_{\Omega_0} \frac{ d\,\Omega'}{\sqrt{ \Omega' - 1 + \exp\Lb - \Omega'\Rb}}\,\,=\,\,z
\eeq
In \eq{BKS5} $\Omega_0$ is the initial condition for $\Omega\Lb z \Rb$: $\Omega\Lb z =0\Rb  = \Omega_0$. Note, that $N_0 $ in \eq{BKS2} is equal to $N_0 = 1 - \exp\Lb \Omega_0\Rb \,\xrightarrow{\Omega_0 \ll 1} \Omega_0$. \eq{BKS5} has two analytic solutions  for small  and large $z$. Indeed, for $ z \,\ll\,1$ $\Omega\Lb z \Rb$ in \eq{BKS5} is close to $\Omega_0$ and for small $\Omega_0$ \eq{BKS5} can be written as follows:
\beq \label{BKS6}
2\intl^{\Omega\Lb z \Rb}_{\Omega_0} \frac{ d\,\Omega'}{\Omega' }\,\,=\,\,z
\eeq 
leading to $\Omega\Lb z \Rb =\Omega_0 e^{\h \,z}$. Hence, the solution of \eq{BKS5} satisfies the initial conditions of \eq{BKS2} at small values of $z$.

For $z \,\,\gg\,\,1$ we can consider $\Omega'$ to be large and approximate \eq{BKS5}
by 
\beq \label{BKS7}
\sqrt{2}\intl^{\Omega\Lb z \Rb}_{\Omega_0} \frac{ d\,\Omega'}{\sqrt{\Omega' }}\,\,=\,\,z
\eeq
which gives $\Omega\Lb z\Rb = \frac{1}{8} z^2$ in accord  with \eq{BKA3}.
 %%%%%%%%%%%%%%%%%%%%%%%%%%%%%%%%%%%%%%%%%%%%%%%%%%%%%%%%%%
 
 \begin{figure}
 	\begin{center}
 	\leavevmode
	\begin{tabular}{c c}
 		\includegraphics[width=8.3cm]{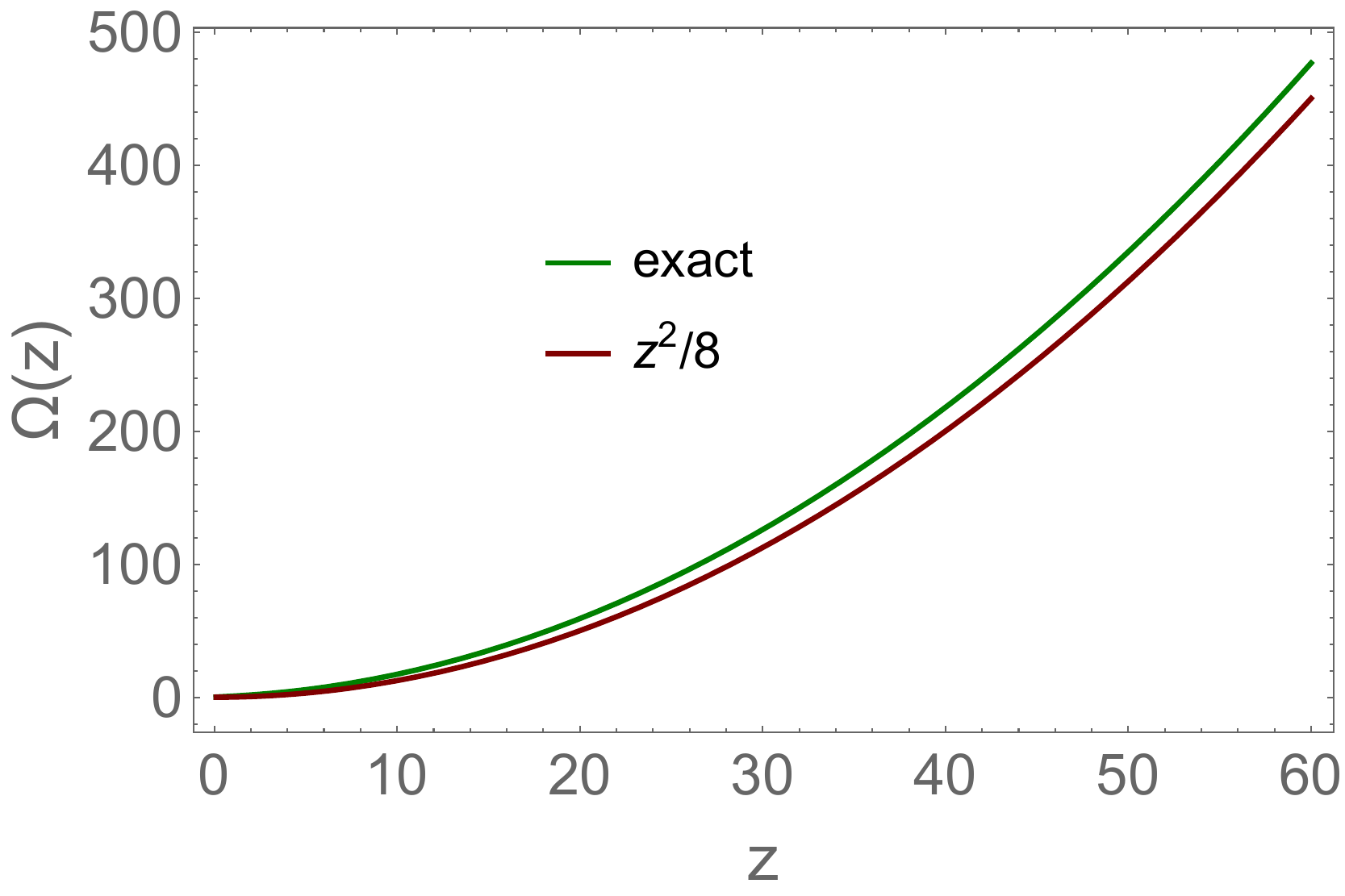}&\includegraphics[width=8cm]{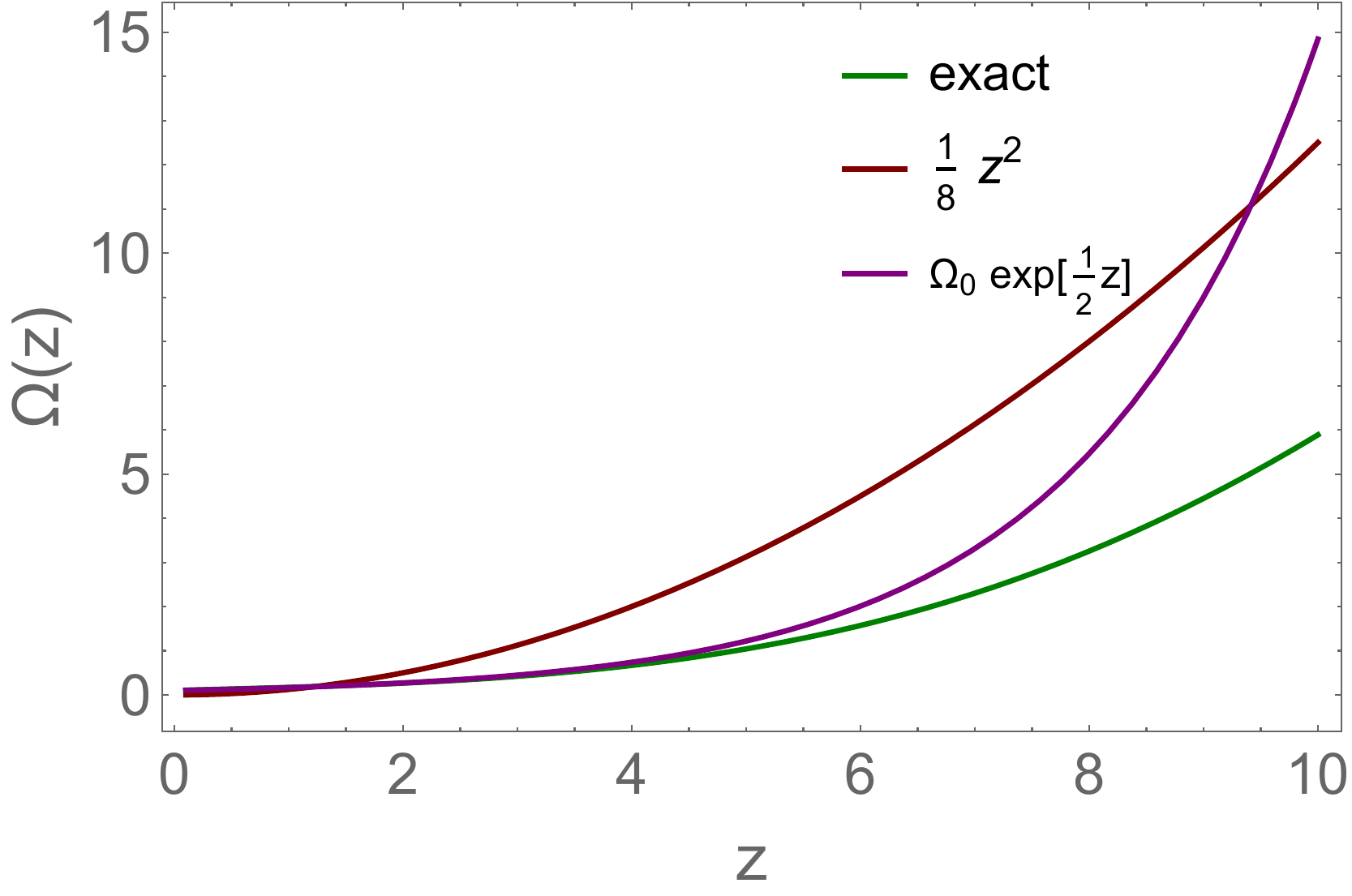} \\
		\fig{om}-a & \fig{om}-b\\
		\end{tabular}	\end{center}
	\caption{ $\Omega\Lb z\Rb$ versus $z$. $\Omega_{exact}$ is the numerical solution of \eq{BKS5} with 
$\Omega_0 =0.1$. \fig{om}-a shows the large range of $z$, while \fig{om}-b is concentrated at small $z< 10$ to illustrate how the exchange of the BFKL Pomeron describes $\Omega\Lb z\Rb$. 
}
\label{om}
\end{figure}
 %%%%%%%%%%%%%%%%%%%%%%%%%%%%%%%%%%%%%%%%%%%%%%%%%%%%%%%%%% 
 
 \fig{om} shows how these analytical limits describe the exact $\Omega\Lb z \Rb$. One can see that correction to the asymptotic behaviour at large $z$ is rather large leading to  $  \frac{1}{8} z^2 - \Omega_{exact} \Lb z\Rb \approx z$. 
 
 The main goal of this paper to obtain information on the dipole densities ($\rho_n $ of \eq{MPSI}) from \eq{BKS1}. 

~

~
        ~
      
                        %%%%%%%%%%%%%%%%%%%%%%%%%%%%%%%%%%%%%%%%%%%%%%%%%%%%  
     \begin{boldmath}
     \section{ Dipole densities and the BFKL Pomeron calculus}
     
     \subsection{ Pomeron solution to QCD evolution equations for $\rho_n$}
      \end{boldmath}

      %%%%%%%%%%%%%%%%%%%%%%%%%%%%%%%%%%%%%%%%%%%%%%%%%%%% 
      
      The evolution equations for $\rho_n$ have been derived in QCD for the BFKL cascade in Ref.\cite{LELU}, and 
     for  $\bar{\rho}_n(\rv, \bv; \{r_i, b_i\})$
defined as
 \beq \label{BRHO}
\bar{\rho}_n(\rv, \bv; \{r_i, b_i\}) \,=\,\,\prod_{i=1}^n \,r^2_i\,\,\rho_n(\rv, \bv; \{r_i, b_i\})
\eeq 
they have the following form 
\bea \label{PD3}
\frac{\partial \,\bar{\rho}_n(\rv, \bv; \{\rv_i, \bv_i\})}{ 
\,\partial\,Y}\,\,&=&\,\,\sum_{i=1}^n\,
\int\,\frac{d^2\,r'}{2\,\pi}\,
K\Lb \vec{r}',\vec{r}_i - \vec{r'}|\vec{r}_i\Rb\\
&\times&\Bigg\{ \bar\rho_n(\rv, \bv; \{\rv_j,\bv_j\}, \rv',\bv_i - (\rv_i - \rv')/2)
\,+\,\bar\rho_n( \rv, \bv; \{\rv_j,\bv_j\}, \rv_i -  \rv',\bv_i - \rv'/2)\,-\,\bar{\rho}_n(\rv, \bv; \{\rv_i, \bv_i\})\Bigg\}
\nn\\
 & & 
\,
+\,\bas\sum_{i=1}^{n-1}\,
\bar\rho_{n-1}(\ldots\,(\vec{r}_i\,+\,\vec{r}_n), b_{in}\dots)~~\mbox{with}~~ K\Lb \vec{r}',\vec{r}_i - \vec{r'}|\vec{r}_i\Rb\,\,\frac{r^2}{r'^2 (\vec{r}_i - \vec{r'})^2}.
\nn
\eea      
  This set of equations  has been derived without assuming that  the BFKL Pomeron calculus is the realization of the Color Glass Condensate (CGC) approach. In this section it will be shown that \eq{PD3}  has a natural solution  in the BFKL Pomeron calculus reproducing the sum of the 'fan' diagrams.
  
  Let us start with $\bar{\rho}_1$, for which we have the following equation:
  
  \beq \label{RHO1}
  \frac{\partial \,\bar{\rho}_1(Y; \rv_1, \bv_1)}{ 
\,\partial\,Y}\,\,=\,\,
\int\,\frac{d^2\,r'}{2\,\pi}\,
K\Lb \vec{r}',\vec{r}_1 - \vec{r'}|\vec{r}_1\Rb
\Bigg\{ \bar\rho_1(Y;  \rv',\bv_1- (\rv_1- \rv')/2)
\,+\,\bar\rho_1(Y; \rv_1 -  \rv',\bv_1 - \rv'/2)\,-\,\bar{\rho}_1(Y; \rv_1, \bv_1)\Bigg\}    
 \eeq

\begin{figure}[ht]
    \centering
  \leavevmode
      \includegraphics[width=10cm]{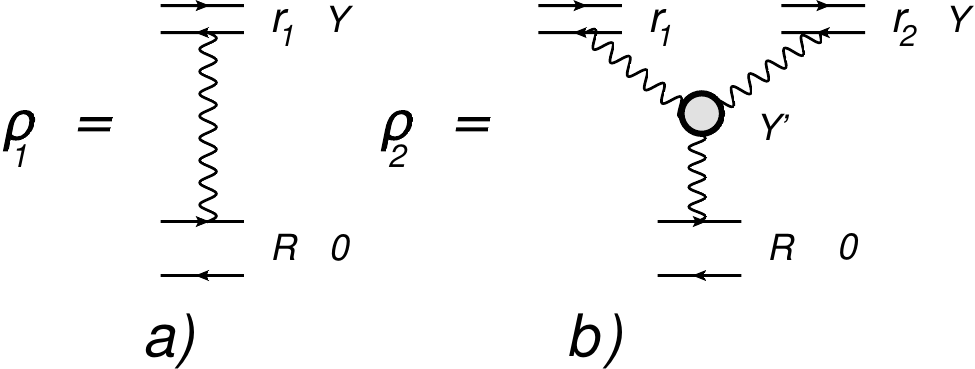}  
      \caption{The dipole densities  in the BFKL Pomeron calculus. \fig{rho}-a shows that $\rho_1$ is given by the exchange of the BFKL Pomeron. \fig{rho}-b demonstrates that $\rho_2$ is the contribution of the first `fan' diagram of the BFKL Pomeron calculus.
    The wavy lines denote the  BFKL Pomeron exchanges.  The circle  denotes the triple Pomeron vertex, given by the BFKL kernel.  }
\label{rho}
   \end{figure}

 Hence one can see that $\rho_1$ is the solution to the BFKL equation and, therefore, describes the BFKL Pomeron exchange (see \fig{rho}-a).
     Using \eq{K2} for the leading twist BFKL kernel and considering $b$ being larger that $r_1$ and $r'$  one can see that \eq{RHO1} is the BFKL linear equation (see \eq{BKS2}).     
      
           %%%%%%%%%%%%%%%%%%%%%%%%%%%%%%%%%%%%%%%%%
%%%%%%%%%%%%%%%%%%%%%%%%%%%%%%%%%%%%%%%%%%
     Our the next step is to show that the first fan diagram of \fig{rho}-b gives the solution of \eq{PD3} for $\bar{\rho}_2$. The contribution of this diagram has the following form (see \fig{rho}-b):
 \bea \label{RHO2}
 &&\bar{\rho}_2\Lb Y; \rv_1,\bv_1, \rv_2, \bv_2\Rb= \\
 && \intl^Y_0 d Y' \int G_{\pom}\Lb Y - Y',\xi_{r_1,r'_1}\Rb\, G_{\pom}\Lb Y - Y',\xi_{r_2,r'_2}\Rb d^2 r'_1 \underbrace{K\Lb \rv'_1,\rv'_2|\rv'_1+\rv'_2\Rb}_{\mbox{triple\,Pomeron\,vertex}}  \frac{ d^2 (\rv_1' + \rv_2')}{(\rv_1' + \rv_2')^2} G_{\pom}\Lb Y', \xi_{\rv_1' + \rv_2', r'}\Rb\nn
 \eea    
   Taking derivative with respect to $Y$ we get:
    \bea \label{RHO21}
 &&\frac{\partial\,\bar{\rho}_2\Lb Y; \rv_1,\bv_1, \rv_2, \bv_2\Rb}{\partial\,Y}= \\
 && \intl^Y_0 d Y'\int  \frac{\partial\,G_{\pom}\Lb Y - Y',\xi_{r_1,r'_1}\Rb}{ \partial \,Y} \, G_{\pom}\Lb Y - Y',\xi_{r_2,r'_2}\Rb d^2 r'_1 K\Lb \rv'_1,\rv'_2|\rv'_1+\rv'_2\Rb  \frac{ d^2 (\rv_1' + \rv_2')}{(\rv_1' + \rv_2')^2} G_{\pom}\Lb Y', \xi_{\rv_1' + \rv_2', r'}\Rb\nn\\
 &&+ \intl^Y_0 d Y' \int  G_{\pom}\Lb Y - Y',\xi_{r_1,r'_1}\Rb \,  \frac{\partial\,G_{\pom}\Lb Y - Y',\xi_{r_2,r'_2}\Rb}{\partial\,Y}  d^2 r'_1\,K\Lb \rv'_1,\rv'_2|\rv'_1+\rv'_2\Rb \frac{ d^2 (\rv_1' + \rv_2')}{(\rv_1' + \rv_2')^2} G_{\pom}\Lb Y',\xi_{\rv_1' + \rv_2', r'}\Rb\nn\\
  &&+ \int G_{\pom}\Lb 0,\xi_{r_1,r'_1}\Rb\, G_{\pom}\Lb 0,\xi_{r_2,r'_2}\Rb d^2 r'_1 K\Lb \rv'_1,\rv'_2|\rv'_1+\rv'_2\Rb  \frac{ d^2 (\rv_1' + \rv_2')}{(\rv_1' + \rv_2')^2} G_{\pom}\Lb Y,\xi_{\rv_1' + \rv_2', r'}\Rb\nn  
    \eea    
    
    The first two terms of this equations    reproduce the first two terms of \eq{PD3} for $\rho_2$ since $G_{\pom}\Lb Y - Y',\xi_{r_1,r'_1}\Rb$ as well as $G_{\pom}\Lb Y - Y', \xi_{r_2,r'_2}\Rb$ satisfy the BFKL equations.    
       
Using the initial conditions of \eq{ZSR} for $\rho_1\Lb Y=0,r_1,b_1\Rb= G_{\pom}\Lb 0, ,\xi_{r_1,r'_1}\Rb$ ( $ \rho_1\Lb Y=0,r_2,b_2\Rb= G_{\pom}\Lb 0, \xi_{r_2,r'_2}\Rb$ )   one can see that the last term in \eq{RHO21} is equal to $\rho_1\Lb Y, \rv_1 + \rv_2, \bv\Rb$. Therefore, \eq{RHO2} gives the solution to \eq{PD3} for $\rho_2$.

           %%%%%%%%%%%%%%%%%%%%%%%%%%%%%%%%%%%%%%%%%
     \begin{figure}[ht]
    \centering
  \leavevmode
      \includegraphics[width=14cm]{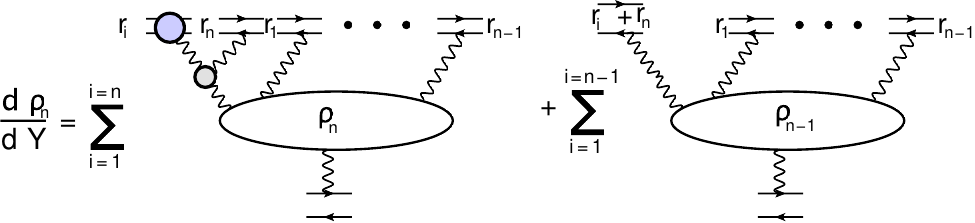}  
      \caption{The graphic form of the equation for $\rho_n$  in the BFKL Pomeron calculus.  in the BFKL Pomeron calculus.
    The wavy lines denote the  BFKL Pomeron exchanges.  The black circle  denotes the triple Pomeron vertex, given by the BFKL kernel. The blue circles stand for the BFKL kernels }
\label{rhon}
   \end{figure}
%%%%%%%%%%%%%%%%%%%%%%%%%%%%%%%%%%%%%%%%%%
Actually, this proof shows a general pattern: the derivatives with respect of Y  of $G_{\pom}^n$ lead to the first term in \eq{PD3} for $\rho_2$, while $Y'=Y$ gives the second term in \eq{PD3} due to the initial conditions. The graphic form of this general derivation one can see in \fig{rhon}.

~

                        %%%%%%%%%%%%%%%%%%%%%%%%%%%%%%%%%%%%%%%%%%%%%%%%%%%%  
     \begin{boldmath}
  \subsection{ Contributions of $\rho_n$ to t-channel unitarity}
      \end{boldmath}

      %%%%%%%%%%%%%%%%%%%%%%%%%%%%%%%%%%%%%%%%%%%%%%%%%%%% 
    In this section we wish to show, that actually only the first term in  the evolution equations for $\rho_n$ ( see \eq{PD3}), contribute to the $t$-channel unitarity of \eq{MPSI}. This term gives the large Pomeron loop diagrams, while the second term leads to the corrections which we cannot control.   
    
    Let us calculate \eq{RHO2} for the leading twist BFKL kernel using \eq{BKS2} for $G_{\pom}$ and \eq{K2} for the kernel. In addition we take into account the initial condition of \eq{ZSR}. Integration over $Y'$ results in the following expression:
    
    \bea \label{RHO22}
 &&\bar{\rho}_2\Lb Y; \rv_1,\bv_1, \rv_2, \bv_2\Rb= \\
 &&\hspace{0.3cm} \h \int G_{\pom}\Lb Y',\xi_{r_1,r'_1}\Rb\, G_{\pom}\Lb Y ,\xi_{r_2,r'_2}\Rb    d^2 r'_1 K\Lb \rv'_1,\rv'_2|\rv'_1+\rv'_2\Rb  \frac{ d^2 (\rv_1' + \rv_2')}{(\rv_1' + \rv_2')^2} G_{\pom}\Lb 0, \xi_{\rv_1' + \rv_2', r'}\Rb\nn\\
 &&- \h\int  G_{\pom}\Lb 0,\xi_{r_1,r'_1}\Rb\, G_{\pom}\Lb 0,\xi_{r_2,r'_2}\Rb  d^2 r'_1 K\Lb \rv'_1,\rv'_2|\rv'_1+\rv'_2\Rb  \frac{ d^2 (\rv_1' + \rv_2')}{(\rv_1' + \rv_2')^2} G_{\pom}\Lb Y, \xi_{\rv_1' + \rv_2', r'}\Rb \nn\eea       
  Using the initial conditions of \eq{ZSR} one obtain:
      \bea \label{RHO23}
\bar{\rho}_2\Lb Y; \rv_1,\bv_1, \rv_2, \bv_2\Rb&= &\h \intl^{\xi_{r_1, r'}}_0 d \xi'  G_{\pom}\Lb Y,\xi_{r_1, r'}-\xi'\Rb\, G_{\pom}\Lb Y,\xi_{r_2, r'}\Rb - \h \int^{\xi_{\rv'_1 +\rv_2, r'}} d \xi'  G_{\pom}\Lb Y,\xi'\Rb\nn\\
&=&  G_{\pom}\Lb Y,\xi_{r_1,R}\Rb\, G_{\pom}\Lb Y,\xi_{r_2, r'}\Rb -  G_{\pom}\Lb Y,\xi_{\rv_1 + \rv_2, r'}\Rb 
   \eea 
 In the second line of \eq{RHO23} we used that $G_{\pom}$ has the form of \eq{BKS2}.

\begin{figure}[ht]
    \centering
  \leavevmode
      \includegraphics[width=10cm]{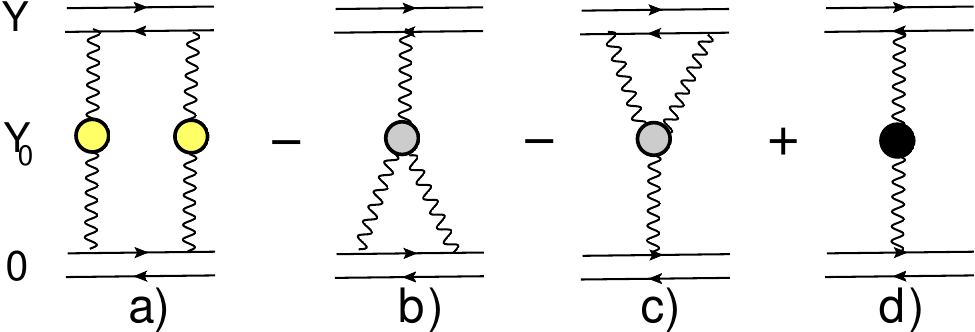}  
      \caption{The graphic form of the  contributions of $\rho_2$ to the scattering amplitude (see \eq{MPSI2}).
    The wavy lines denote the  BFKL Pomeron exchanges.  The gray circles  denote the triple Pomeron vertex, given by the BFKL kernel. The yellow circles corresponds to integration over $\gamma^{BA}$  in \eq{MPSI2}. The black circle stands for the contribution to the intercept of the BFKL Pomeron of the order of $\bas^2$. }
\label{rho2a}
   \end{figure}

   The contributions of $\rho_2$ to the t-channel unitarity for the scattering amplitude, given by    \eq{MPSI}, have the form:
   \bea \label{MPSI2}
     && N_2\Lb Y, r, r' ;  \vec{b}\Rb\,= -\,2\int  d^2 r_1\,d^2\,r'_2\,d^2 b'_1,d^2 b'_2 \,d^2 \delta b_1,d^2 \delta b_2\\
     && \gamma^{BA}\Lb r_1,r'_1, \vec{b}_1-  \vec{b'_1}\equiv \delta \vec{b} _1\Rb 
  \gamma^{BA}\Lb r_2,r'_2, \vec{b}_2-  \vec{b'_2}\equiv \delta \vec{b} _2\Rb     \,\,\rho_2\Lb Y - Y_0,  \vec{r}_1, \vec{r}_2,\vec{b}_1,\vec{b}_2\Rb\,\rho_2\Lb Y - Y_0,  \vec{r}'_1, \vec{r}'_2,\vec{b}'_1,\vec{b}'_2\Rb\nn\eea
  Plugging \eq{RHO23} into \eq{MPSI2} and integrating over all $r$'s and $b$'s we obtain the contribution which graphic form is shown in \fig{rho2a}
  
           %%%%%%%%%%%%%%%%%%%%%%%%%%%%%%%%%%%%%%%%%
%%%%%%%%%%%%%%%%%%%%%%%%%%%%%%%%%%%%%%%%%%
   Using the t-channel unitarity for the BFKL Pomeron \cite{MUDI,MUSA,GLR} we have
 \beq \label{RHO24}
   G_{\pom}\Lb z\Rb \,\,=\,\,\int \frac{ d^2 r_1 d^2 r'_1 d^2 b_1 d^2 b'_1}{r^4_1\,r'^4_2}\,G_{\pom}\Lb Y - Y_0, \xi_{r,r_1}\Rb \gamma^{BA}\Lb r_1,r'_1, \vec{b}_1-  \vec{b'_1}\equiv \delta \vec{b} _1\Rb G_{\pom}\Lb Y_0, \xi_{r'_1, r'}\Rb   \eeq   
  where  $z$  is determined by \eq{zz}.
  Bearing this in mind we see that the diagram \fig{rho2a}-a is equal to the exchange of two Pomeons and does not depend on $Y_0$ being proportional to $
  G^2_{\pom}\Lb Y - Y_0, \xi_{r,r_1}\Rb  G^2_{\pom}\Lb Y_0, \xi_{r'_1,  r'}\Rb=  
  G^2_{\pom}\Lb Y , \xi_{r, r'}\Rb\,=\,  \exp\Lb 4\,\bas \,Y  \,+\, \xi_{r, r'}\Rb$ The diagrams \fig{rho2a}-b and \fig{rho2a}-c do depend on $Y_0$ and they are suppressed by  a factor $
  \exp\Lb - 2\bas\,( Y-Y_0)\Rb \Lb\exp\Lb - 2 \bas\,Y_0\Rb\Rb $ in comparison with \fig{rho2a}-a, respectively.  These diagrams belong to the BFKL Pomeron calculus but  have to be integrated over $Y_0$. { Note, that in \eq{MPSI}  they enter at fixed $Y_0$. Actually, \eq{MPSI} states that the scattering amplitude $ N\Lb Y, r, r' ;  \vec{b}\Rb$ should not depend on the value of $Y_0$. It means that the diagrams of \fig{rho2a}-b and of \fig{rho2a}-c
   have to be viewed as  corrections. As we have shown they are smaller than \fig{rho2a}-a.
\fig{rho2a}-d gives a contribution to the intercept of the Pomeron of the order of   $ \bas^2$. However, since its position is fixed at Y=$Y_0$ this diagram cannot describe the contribution of the next order corrections to the Pomeron intercept in the scattering amplitude. Hence, all diagrams except \fig{rho2a}-a lead to the correction in summing of the large Pomeron loops. In other words, for summing the contributions of the large Pomeron loops to the scattering amplitude we need to account only for the contributions to $\rho_n$ of the following form:
   \beq \label{RHONF} 
   \rho_n\Lb Y - Y_0, \{ \vec{r}_i,\vec{b}_i\}\Rb\,\,=\,\,C_n  \prod_{i=1}^n G_{\pom} \Lb z_i\Rb   \eeq
   where  $z$ is given by \eq{zi} with $\xi_{r_i,r}$  from  \eq{XI} .  
  $r$ is the size of the incoming dipole. $C_n$ is a constant which value could depend on $r$ . 
   
   ~
   
    ~
 
                        %%%%%%%%%%%%%%%%%%%%%%%%%%%%%%%%%%%%%%%%%%%%%%%%%%%%  
     \begin{boldmath}
\section{ BK scattering amplitude in the BFKL Pomeron calculus}
      \end{boldmath}   
      Using \eq{MPSI} and \eq{RHONF} we can calculate dipole-nucleus amplitude  putting $Y_0 =0$ (see \fig{mpsi1}).  For $\rho_n$ of the nucleus we
 consider a nucleus as a bag of dipoles with size $R$ which do not interact with each other at $Y_0  = 0$.
It means that (see \eq{ZSR} and \fig{mpsi1})
\beq \label{SABK1}
\rho^A_n\Lb Y_0=0, r', b',  \{ \vec{r}'_i,\vec{b}_i'\}\Rb \,\,=\,\,\frac{S^n_A\Lb b'\Rb}{n!}\prod^n_{i=1} \delta^{(2)}\Lb \vec{r}_i' - \vec{R}\Rb \delta^{(2)}\Lb \vec{b}' - \vec{b}_i\Rb ;
\eeq
 Factor $S^n_A\Lb b' \Rb$ describes the probability to find $n$- nucleons (dipoles) in a nucleus at the impact parameter $b'$. 
 
 Plugging \eq{SABK1} into \eq{MPSI} we obtain:
 \beq \label{SABK2}
 N^{dip}_{A}\Lb Y, r,R ;  \vec{b}\Rb  = \sum_{n=1}^\infty C_n (-1)^n \prod^n_{i=1}  \int \frac{ d^2 r_i }{r^2_i}\,G_{\pom}\Lb Y, \xi_{r}\Rb \gamma^{BA}_A\Lb r_i, b\Rb G_{\pom}\Lb Y_0, \xi_{r'_1,R}\Rb \eeq
 where $\gamma^{BA}_A\Lb r_i, b\Rb $ is the amplitude of scattering of the dipole with size $r_i$ with the nucleus at small energies.  Using \eq{RHO24} we can rewrite this equation in the following form:
 \beq \label{SABK3}
 N^{dip}_{A}\Lb Y, r,R ;  \vec{b}\Rb  = \sum_{n=1}^\infty C_n (-1)^{n-1} G^n_{\pom}\Lb Y,  \xi_{r,R},b\Rb\eeq
     where $G_{\pom}\Lb Y,  \xi_{r,R},b\Rb   = N_0\exp\Lb \h z_A\Rb$ with
    $z_A = 4 \,\bas\, Y +  \xi_{r,R_A}   =   4 \,\bas\, Y + \ln\Lb r^2 Q_A\Lb Y=0, b\Rb\Rb $.
    $Q_A\Lb Y=0, b\Rb  = Q_0 S_A\Lb b\Rb$.
    
    \eq{SABK2} shows that we can find coefficients $C_n$ in \eq{RHONF} by finding the BK amplitude $N^{dip}_{A}\Lb Y, r,R ;  \vec{b}\Rb$ as a series of many BFKL Pomeron exchange. In other words, we can find $C_n$ from \eq{BKS1} by rewriting it  as the equations for $C_n$.   Let us first retrieve these coefficients from \eq{BKA2} plugging $S_{01}\Lb z_A\Rb = 1\,-\, N^{dip}_{A}\Lb z_A\Rb$ in this equation. One can see that the equations for $C_n$  takes the form:
 \beq \label{SABK30}
    \sum_{n=0}^\infty C_n (-1)^{n}  \frac{\kappa}{2} \,n\, G^n_{\pom}\Lb Y,  \xi_{r,R},b\Rb\,=\,- z_A \sum_{n=0}^\infty C_n (-1)^{n}  \, G^n_{\pom}\Lb Y,  \xi_{r,R},b\Rb    ~~~\mbox{or} ~~~~\langle|n|\rangle =  - \frac{2}{\kappa} \,z_A
    \eeq
     Assuming that we calculate this sum over $n$ using the method of steepest descent, we see that  the saddle point value of $n$: $n_{SP} = -     \frac{2}{\kappa} \,z_A \,<\,0$.
 We think that we can understand this strange result only if we can make the analytic continuation of  \eq{SABK3}  to the complex $n$. In this case we can rewrite $S\Lb z_A\Rb$  as the integral over  contour $C$ in \fig{cont}-a, viz. 
     \beq \label{SABK31}
 S\Lb z_A\Rb  =\oint_C \frac{d n}{2\,\pi\,i}  C_n \frac{\pi}{\sin\Lb \pi\,n\Rb} G^n_{\pom}\Lb z_A\Rb\eeq   
 Assuming that we can open the contour $C$  to the contour $C'$  we can take the integral over the imaginary  $n$ using the method of steepest descent.  Denoting $C_n =\exp\Lb \phi(n)\Rb$ we obtain the equation for the saddle point $n_{SP}$:
\beq \label{SABK32}
\phi'_n \Lb i\,n_{SP}\Rb=i\,n_{SP}= - \frac{2}{\kappa} \,z_A
\eeq
 From this equation we find $\phi\Lb n \Rb = \frac{1}{2} n^2 $  for $\kappa = 4$  and the integral over imaginary  $n$ ( contour C') gives the expected result:
    \beq \label{SABK33}
 S\Lb z_A\Rb  =  C \exp\Lb - \frac{z^2_A}{8}\Rb
 \eeq  
 
 The constant $C$ has to be found from  matching with the solution at small $z_A$.
 Actually, as it has been discussed in the introduction, a smooth function $C(z)$ also 
 leads  to the solution of the BK equation at large $z_A$.
 
 Therefore, we found out that coefficients $C_n = \exp\Lb \frac{1}{2}\,n^2\Rb$.  The asymptotic Pomeron series of \eq{SABK3} cannot be summed using the Borel procedure \cite{BORSUM}. However, such behaviour of $C_n$  is not new and, in particular, the one dimensional model  of Ref.\cite{MUSA},  has such Pomeron series, which could be summed (see Ref.\cite{KLLN} appendix A). We will discuss in detail the procedure of summation but we wish to state  that the procedure, suggested in Ref.\cite{KLLN},  has been checked in the model. It reproduces correct scattering amplitude 
 and allow us to make the analytic continuation that we have used.
 
 Now we come back to the nonlinear equation (see \eq{BKS1}),  which we rewrite for the scattering matrix:
 \beq \label{SABK34}
 \kappa\frac{ d \,S\Lb z_A\Rb}{d \,z_A}\,\,=\,\, -z_A \,S\Lb z_A\Rb\,\,+\,\,S\Lb z_A\Rb\intl^{z_A}_0 d z' S\Lb z'\Rb
\eeq  
Recall that $\kappa = 4$ for the leading twist kernel.  Rewriting this equation in the form:
 \beq \label{SABK35}
 \kappa\frac{ d \,S\Lb z_A\Rb}{d \,z_A}\,\,=\,\, -\,\Lb z_A \,-\,z_0\Rb\,S\Lb z_A\Rb\,\,-\,\,S\Lb z_A\Rb\intl^\infty_{z_A} d z' S\Lb z'\Rb~~~\mbox{with}~~~z_0 \,=\,\intl^\infty_{0} d z' S\Lb z'\Rb\eeq 
 Plugging in this equation the solution:
  \beq \label{SABK36} 
 S\Lb z_A\Rb\,=\,C\Lb z_A\Rb \exp\Lb - \frac{\Lb z\,-\,z_0\Rb^2}{2\,\kappa}\Rb
 \eeq
 one can see that for $C\Lb z_A\Rb \,< \, \exp\Lb - \frac{\Lb z\,-\,z_0\Rb^2}{2\,\kappa}\Rb$  the nonlinear term gives a small contribution in the kinematic region where $  \exp\Lb - \frac{\Lb z\,-\,z_0\Rb^2}{2\,\kappa}\Rb\,\ll\,1$.

                %%%%%%%%%%%%%%%%%%%%%%%%%%%%%%%%%%%%%%%%%
     \begin{figure}[ht]
    \centering
  \leavevmode
  \begin{tabular}{c c c}
      \includegraphics[width=6cm]{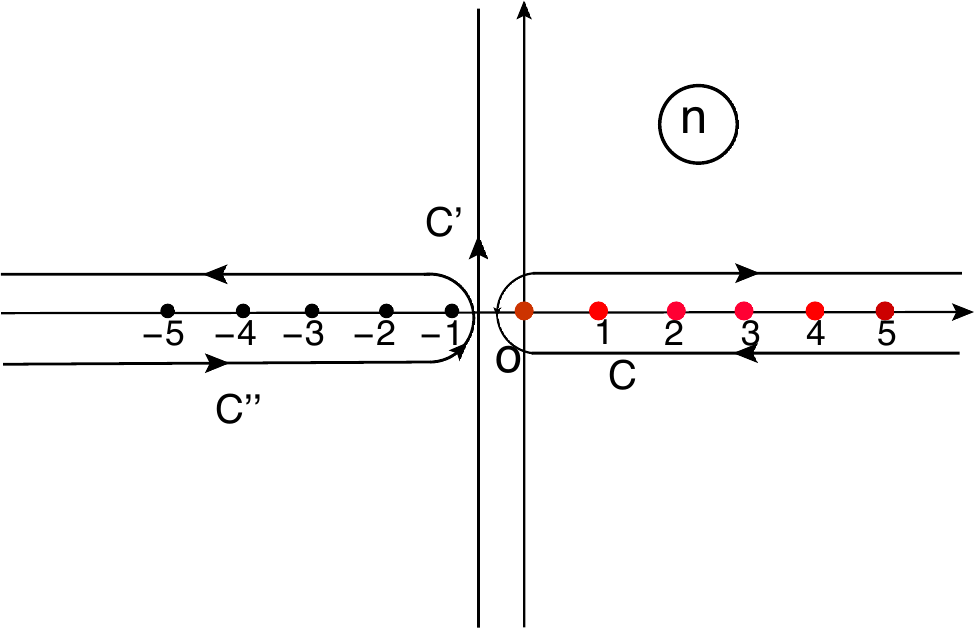} & & \includegraphics[width=6cm]{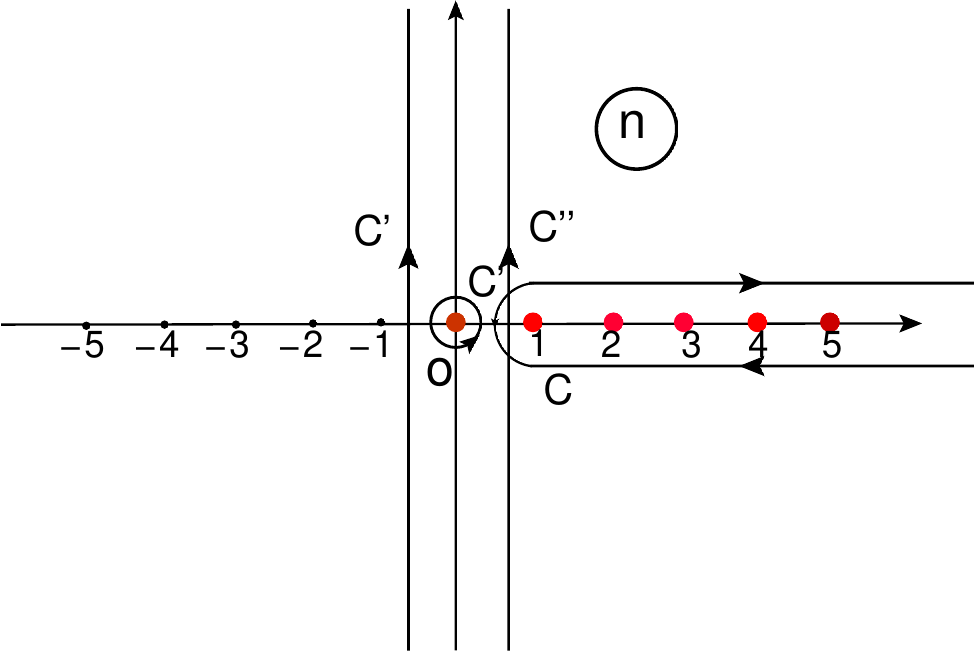} \\
      \fig{cont}-a & ~~~~~& \fig{cont}-b\\
      \end{tabular}
           \caption{The contours of integration in \eq{SABK31}(\fig{cont}-a) and in \eq{SABK3}(\fig{cont}-b). The red points denote the poles in $n$. The black points show the poles in the left semi-plane where  the scattering matrix for dipole-dipole interactions ( $S^{d-d}\Lb z\Rb$) has the double poles.}
\label{cont}
   \end{figure}
%%%%%%%%%%%%%%%%%%%%%%%%%%%%%%%%%%%%%%%%%%     
  Generally  to specify function $C\Lb z_A\Rb$ in \eq{SABK36}  we    plug \eq{SABK3} into \eq{BKS1} and  have the following equations for $C_n$:
 \beq \label{SABK4}
    2\,n C_n=  \frac{2}{n} C_n \,-\,2\sum_{k=1} ^n \frac{C_k\,C_{n-k}}{k}\,\,=\,\,
 \frac{2}{n} C_n\,+\,\,n \sum_{k=1} ^n \frac{C_k\,C_{n-k}}{k\, (n-k)}
 \eeq
  Denoting $C_n/n $ by  $ c_n $ we obtain:
   \beq \label{SABK5}
    2\,\Lb n^2\,-\,1\Rb c_n\,= \,\,n \sum_{k=1} ^n c_k c_{n-k} \eeq        
    The recurrent relation has the initial condition $C_1=1$, which give $c_2=1/3, c_3 =1/8$. At large $n$    \eq{SABK5} has solution: $c_n = 2 e^{\alpha\,n}$ with arbitrary $\alpha$. However, this solution does not reproduces the large $z$ asymptotic behaviour of \eq{I1}\cite{LENP}.  As we have discussed above we  expect $C_n \propto \exp\Lb \h \,n^2\Rb$. To show how such dependence of $C_n$ appears in \eq{SABK5} we  first rewrite \eq{SABK3} as the integral over $n$. It takes the form (see \fig{cont}-b):
    \beq \label{SABK6}
 N^{dip}_{A}\Lb Y, r,R ;  \vec{b}\Rb  =\oint_C \frac{d n}{2\,\pi\,i}  C_n \frac{\pi}{\sin\Lb \pi\,n\Rb} G^n_{\pom}\Lb Y,  \xi_{r,R},b\Rb\eeq
    
       Bearing in mind \eq{SABK6} we can view $n$ and $k$ as   continuous variables, replacing \eq{SABK5} as follows:
    \beq \label{SABK7}
    2\,\Lb n^2\,-\,1\Rb c_n\,= \,\,n \intl^{n-1}_1 \!\!\!d k \, c_k c_{n-k} \eeq            
  We are looking for the solution in the form: $c_n\,\,= \,\,\phi\Lb n\Rb \exp\Lb \h \, n^2\Rb$ where $\phi\Lb n \Rb$  is a smooth function of $n$ in comparison with $ \exp\Lb \h\, n^2\Rb$. With such $c_n$ the integral over $k$ in \eq{SABK7} can be taken using the method of steepest decent, since the integrant has  strong maxima at the limits of integrations: $k =1$ and $k =n-1$. Introducing a new variable $k = 1+\delta k $ in the first maximum, one can see that the main contribution comes from small $\delta k \sim \frac{1}{n}$ at large $n$. Rewriting $\h\Lb (n - k)^2 + k^2\Rb $   as $ \kappa(\h (n-1)^2  - n \delta k + \delta k^2)$ one can see that
the equation takes the form at large $n$:
   \beq \label{SABK8}
    2\,n^2\, \phi(n)\,= \,\,2\, \phi\Lb n - \frac{1}{n}\Rb \phi\Lb1-\frac{1}{n}\Rb e^{ - n + 2n} \eeq       
     with the solution
       \beq \label{SABK90}   
       \phi\Lb 1 -\frac{1}{n}\Rb= n^2\,\,e^{n};~~~~~~~\phi\Lb x\Rb\,\,=\,\,\frac{1}{(1-x)^2}\exp\Lb\frac{1}{x-1}\Rb
       \eeq
       Finally for large $n$ we have
          \beq \label{SABK9} 
         C_n =    \exp\Lb \h\, n^2\,\, -  \,\, 2\,n\Rb   
         \eeq
         Plugging this solution to the scattering amplitude of \eq{SABK6} we obtain:
             
         \bea \label{SABK10}
 &&N^{dip}_{A}\Lb Y, r,R ;  \vec{b}\Rb =\\
 &&\oint_C\frac{ d n}{2\,\pi\,i} \frac{1}{n} \exp\Lb \h n^2 -   2\,n\Rb    \frac{\pi}{\sin\Lb \pi\,n\Rb} G^n_{\pom}\Lb Y,  \xi_{r,R},b\Rb=
 \oint_C\frac{ d n}{2\,\pi\,i}  \frac{1}{n} \exp\Lb \h \,n^2 -   2\,n\Rb    \frac{\pi}{\sin\Lb \pi\,n\Rb} e^{\h \,n\,z_A}\nn
 \eea
    This equation is a good illustration of a general fact that
the multi-Pomeron expansion appears as a badly divergent series. Summation of  asymptotic series implies finding an  analytic function with identical series expansion.  In general, this asymptotic series cannot be summed 
even via Borel resummation\cite{BORSUM}. Fortunately, Ref.\cite{KLLN} has solved the main difficult problems that we are faced with the resummation of the asymptotic series 
of \eq{SABK10} type. The first problem is that $C_n \propto \exp\Lb \h n^2\Rb$ which makes \eq{SABK10} not Borel summerable. Plugging in \eq{SABK10} the integral representation for $C_n = \exp\Lb \h n^2\Rb$:
\beq \label{SABK11} 
\exp\Lb \h n^2\Rb= \frac{1}{\sqrt{2 \pi}}\intl^{\infty}_{-\infty} d \lambda \exp\Lb \,- \h \lambda^2 \,- n\,\,\lambda\Rb     
         \eeq
one can see that the asymptotic series at fixed $\lambda$ can by summed {\it a la} Borel\cite{BORSUM}  and can be extended  by analytic continuation  to any value of $n$ in the right half-plane.  Therefore, we can integrate   \eq{SABK10}  along the contour $C'$ in \fig{cont}-b. Using the Borel prescription we rewrite this equation in the following form:
 \begin{subequations}
 \bea \label{SABK12}
 N^{dip}_{A}\Lb Y, r,R ;  \vec{b}\Rb
&= & \oint_{C'}\frac{ d n}{2\,\pi\,i}\intl^{\infty}_0 d t e^{-t}  \frac{1}{n} \exp\Lb \h\,n^2 -   2\,n\Rb    \frac{\pi}{\sin\Lb \pi\,n\Rb\,\Gamma\Lb n+1\Rb} \,t^n\,e^{\h \,n\,z_A}\ \label{SABK12a}\\
&=& \oint_{C'}\frac{ d n}{2\,\pi\,i}\intl^{\infty}_0 d t e^{-t}  \frac{1}{n} \exp\Lb \h \,n^2 -   2\,n\Rb    \Gamma\Lb - n\Rb  \,t^n\,e^{\h \,n\,z_A} \label{SABK12b}\\
&=& 1 + \intl^{-\epsilon + i \infty}_{- \epsilon - i \infty} \frac{ d n}{2\,\pi\,i}\intl^{\infty}_0 d t e^{-t}  \frac{1}{ n} \exp\Lb \h \,n^2 -   2\,n\Rb    \Gamma\Lb - n\Rb  \,t^n\,e^{\h \,n\,z_A} \label{SABK12c}
\eea
\end{subequations}
In \eq{SABK12a} and \eq{SABK12b} the formulae { \bf 8.332(2), 8.334}  {of Ref.\cite{RY} are used. In \eq{SABK12c} we integrated over the pole at $n=0$. Note that we assumed that $C_{n=0}$ is equal to 1 to reproduce the unitarity limit $N^{dip} \,\to\,1$ at $z \to \infty$.   \eq{SABK9} cannot provide us with the analytic continuation of $C_n$ at small values of $n$.  Fortunately, we do not need to know $C_n$ at small $n$.

Indeed, taking the integral in \eq{SABK12c} using the steepest decent method we see that $  n_{SP} = i\,\h \,z_A$.  Therefore, the large value of $n$ are essential in this integral. Bearing this in mind we can safely use $C_n$ at large $n$ for calculating this integral. Hence
the scattering amplitude is equal to
 \beq \label{SABK13}
 N^{dip}_{A}\Lb Y, r,R ;  \vec{b}\Rb\,\,=\,\,1\,\,\,-\,\,\exp\Lb - \frac{1}{8} z^2_A + z_A \,-\,\ln\Lb \h z_A\Rb  - 2\Rb
 \eeq
    Therefore, we reproduce \eq{I1}  and fixed the value of the smooth function $C(z)$.
    For the phase $\Omega\Lb z_A\Rb $ we have
    \beq \label{SABK14}
    \Omega\Lb z_A\Rb     \,=\,\, \frac{1}{8} z^2_A - z_A \,+\,\ln\Lb \h z_A\Rb \,+\,2
    \eeq
    
    In \fig{omn} we compare this solution with the exact numerical solution of \eq{SABK5}. One can see that we found  a good approximation for  $z \geq 4$.      
    %%%%%%%%%%%%%%%%%%%%%%%%%%%%%%%%%%%%%%%%%%%%%%%%%%%%%%%%%%
\begin{figure}
 	\begin{center}
 	\leavevmode
		\begin{tabular}{l l} 
 		\includegraphics[width=8.8cm]{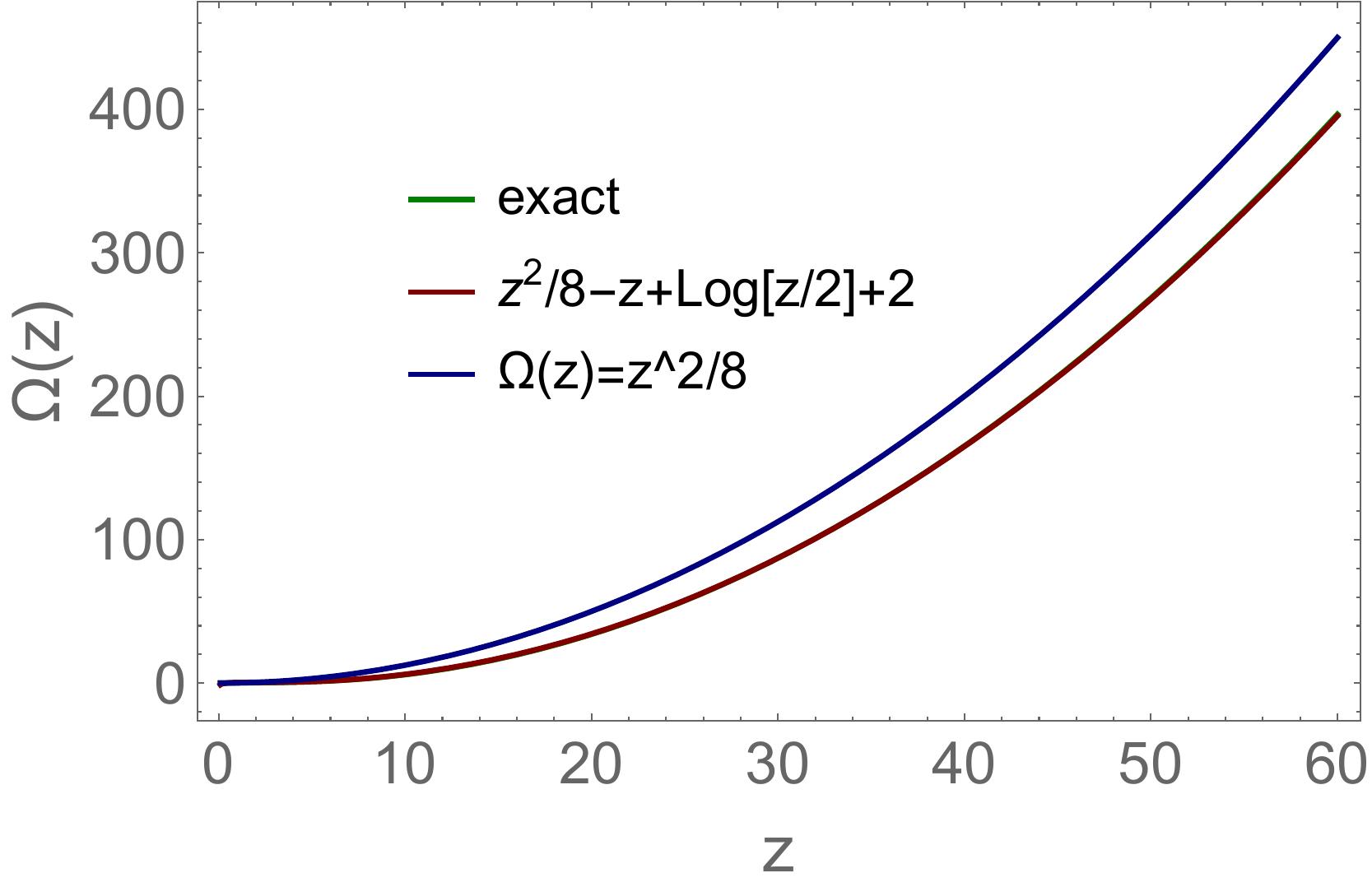}&\includegraphics[width=8.45cm]{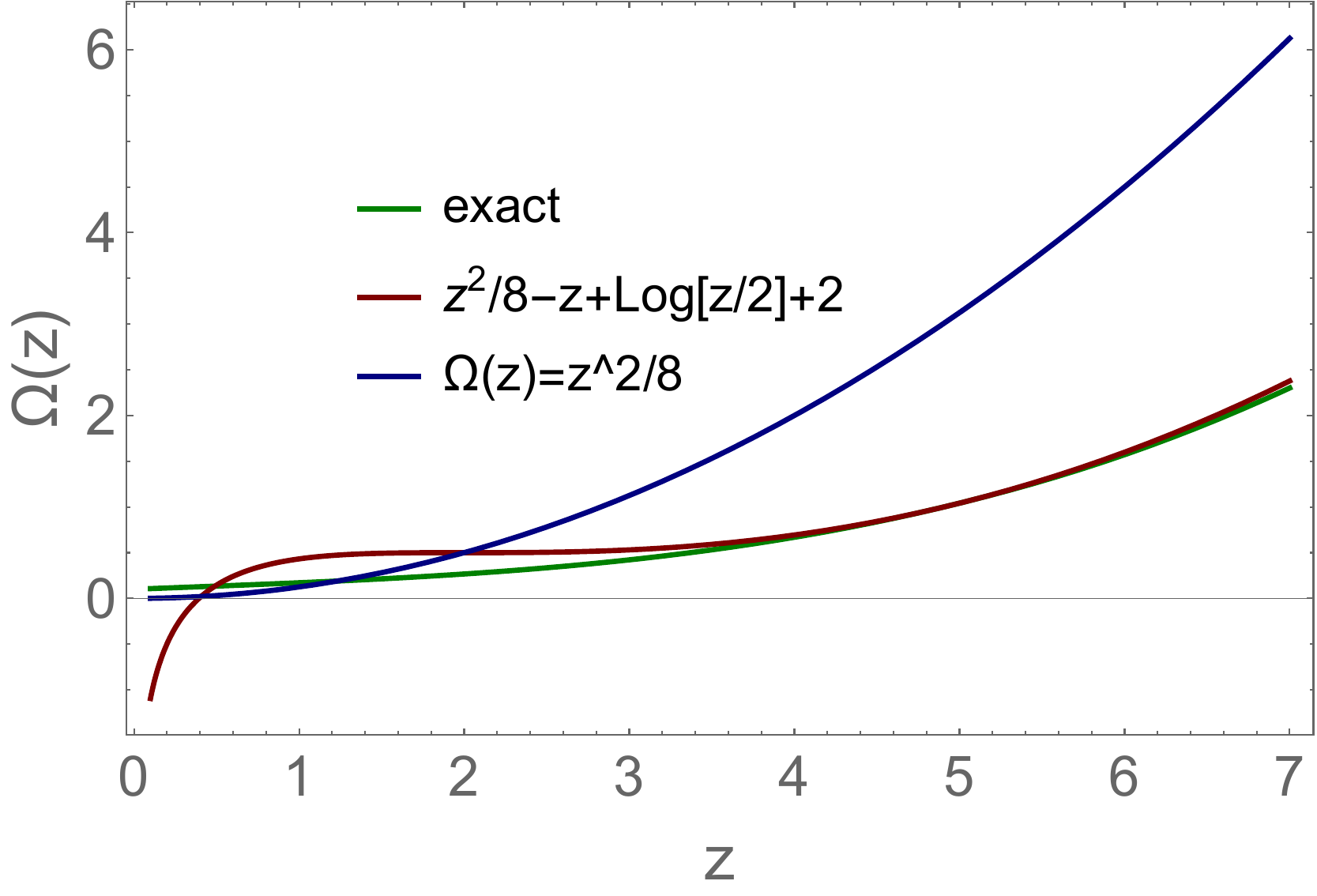} \\
\fig{omn}-a & \fig{omn}-b\\
\end{tabular}	
		\end{center} 
		\caption{ $\Omega\Lb z\Rb$ versus $z$. $\Omega_{exact}$ is the numerical solution of \eq{BKS5} with 
$\Omega_0 =0.1$. $\Omega\Lb z\Rb    \,=\,\, \frac{1}{8} z^2_A - z_A \,+\,\ln\Lb \h z_A\Rb$ is our approximation for $\Omega$ that is coming from the large $n$ behaviour of the coefficient $C_n$ in the Pomeron expansion of \eq{SABK15}.  $\Omega = z^2/8$ is the asymptotic solution of Ref.\cite{LETU}.
}
\label{omn}
\end{figure}
 %%%%%%%%%%%%%%%%%%%%%%%%%%%%%%%%%%%%%%%%%%%%%%%%%%%%%%%%%%     
      Our main problem that we found the solution to the master equation (see \eq{SABK4}) only in the region of large $n$. Hence, we cannot use the initial condition $C_1=1$ to determine all coefficients. In particularly  the general form of solution:
\beq \label{SABK15} 
         C_n =   \frac{1}{n} \exp\Lb \beta  \h n^2\,\, -  \,\, 2\alpha\,n\Rb   
         \eeq
  with arbitrary coefficients $\beta$ and $\alpha$ is also the solution.  We use the behaviour  of the scattering amplitude at large $z$ ( the solution to \eq{BKA2})  to fix $\beta =1$. To find $\alpha$ we need to evaluate the value of $C_n$ at small $n$. Since we do not know this behaviour we use the numerical solution to \eq{BKS5}  to choose the value of $\alpha$.

  Therefore, $\alpha =\beta=1$ is   our preferable choice.   Since, these values lead to good descriptions at small values of $z_A$ (see \fig{omn}-b)  we can hope that our $C_n$ could describe the region of small $n$.   It is instructive to note that $-z_A +2 $ term in \eq{SABK14} comes from the choice $\alpha =1$ while $\ln(z_A/2)$ term reflects $1/n$ behaviour of $C_n$.  Comparing \eq{SABK33} and \eq{SABK36} with our approach with $C_n$ we see that the only advantage in using $C_n$  is  that \eq{SABK15} allows us to find the value of the dipole density using \eq{RHONF} and generalize our approach to dilute-dilute and dense-dense system scattering using \eq{MPSI}.
  
  Concluding this section we wish to point out  that the BK  scattering amplitude of this section  only  has  minor differences with the one discussed in our previous paper \cite{LEDIDI}. However, here and in Ref.\cite{LEDIDI} different  approaches have been used to sum the asymptotic Pomeron series. The advantage of this paper is that the approach for summing the non Borel summarisable series has been checked in the one dimensional model \cite{MUSA,KLLN} in which this method reproduced the  exact  scattering amplitude. One of the check of  the difference between two approaches is the multiplicity distribution which we will consider below. 
      
      ~
      
      ~

                                     %%%%%%%%%%%%%%%%%%%%%%%%%%%%%%%%%%%%%%%%%%%%%%%%%%%%  
     \begin{boldmath}
     \section{Multiplicity distribution and entropy   of produced gluons for BK scattering amplitude }
      \end{boldmath}

      %%%%%%%%%%%%%%%%%%%%%%%%%%%%%%%%%%%%%%%%%%%%%%%%%%%%   
  
         ~

      In this section we are going to make use of the advantage of the multi Pomeron expansion for the scattering amplitude (see  series of \eq{SABK3} ) 
      to obtain the multiplicity distribution of produced gluons for BK scattering amplitude.    
Our approach consists of two steps (see Ref.\cite{MUSA,LEDIDI}). First, we recall that it is proven in Refs.\cite{BFKL} that the $s$-channel unitarity for the BFKL Pomeron has the form:
  \beq \label{MD1}
2\, { \rm Im} \,G_{\pom}\Lb z \Rb\,\,=\,\,\sigma^{\mbox{\tiny BFKL}}_{in}(z)
\eeq
 where $G_{\pom}$ is the Green's function for the BFKL Pomeron.
  $\sigma^{\mbox{\tiny BFKL}}_{in}(z)$ is the inelastic cross sections of  produced gluons , which  have the Poisson distribution  with this mean multiplicity $\bar{n} = \h\,z$ (see Appendix A of Ref.\cite{LEDIDI}).
 
 The second step is the AGK cutting rules\cite{AGK}, which allow us to calculate the imaginary part of   the scattering amplitude, that determines the cross sections, through the powers of ${ \rm Im} \,G^{\mbox{\tiny BFKL}}\Lb z\Rb$\footnote{In the widespread slang this contribution is called by   cut Pomeron.}. Our master formula takes the form of convolution for the cross section of produced $n$  gluons:
 \beq \label{MD2}
 \sigma_n\Lb z \Rb\,\,=\,\,\sum_k\underbrace{ \sigma_k^{AGK}\Lb z \Rb}_{ \propto\,\Lb{\rm Im} G_{\pom} \Rb^k}\underbrace{ \frac{\Lb k \, \h \,z\Rb^n}{n!} e^{ -k\, \h\,z}}_{\mbox{Poisson distribution}}
\eeq

The AGK cutting rules \cite{AGK} allows us to calculate the contributions of $n$-cut Pomerons if we know $F_k$: the contribution of the exchange of $k$-Pomerons to  the cross section. They take the form:
   \begin{subequations} 
    \bea 
n\,\geq\,1:\sigma^k_n\Lb Y, \xi_{r,R}\Rb&=& (-1)^{n -k}\frac{k!}{(n - k)!\,n!}\,2^{k}\, F_k(Y,\xi_{r,R})\label{AGKK}\\
n\,=\,0:\sigma^k_0\Lb Y\Rb&=&\Lb -1\Rb^k \Bigg(2^k\,\,-\,\,2\Bigg) F_k(Y,\xi_{r,R});\label{AGK0}\\
\sigma_{tot}&=&\,\,2 \sum_{k=1}^\infty (-1)^{k+1} \,F_k(Y,\xi_{r,R});\label{XS}\,
\eea
 \end{subequations} 
$\sigma_{tot} = 2\, {\rm Im} A\Lb z\Rb$ where $A$ is the scattering amplitude. $\sigma_0$ is the cross section of diffractive  production of small numbers of gluons which is much smaller than  $\Delta\, Y$. First we rewrite \eq{SABK3} in more convenient form to find $F_k(Y,\xi_{r,R})$ :
\bea \label{MD3}
&& N^{dip}_{A}\Lb Y, r,R ;  \vec{b}\Rb  = \\
&&\frac{1}{\sqrt{2\,\pi}}\intl^{\infty}_{-\infty}\!\!\!d \lambda \, e^{ - \h \lambda^2}\sum_{n=1}^\infty  (-1)^{n-1}  e^{ \h\,n\, \Lb z_A\,-\,4\,-\,2 \lambda \Rb}\,\,=\,\, \frac{1}{\sqrt{2\,\pi}}\intl^{\infty}_0\!d t \intl^{\infty}_{-\infty}\!\!\!d \lambda \, e^{ - \h \lambda^2}\sum_{n=1}^\infty  (-1)^{n-1}e^{ \h\,n\, \Lb z_A\,-\,4\,-\,2 \lambda\,-2\,t \Rb} \nn
 \eea

    %%%%%%%%%%%%%%%%%%%%%%%%%%%%%%%%%%%%%%%%%%%%%%%%%%%%%%%%%%
 
 \begin{figure}[ht]
 	\begin{center}
 	\leavevmode
	\begin{tabular}{l l} 
 		\includegraphics[width=9cm]{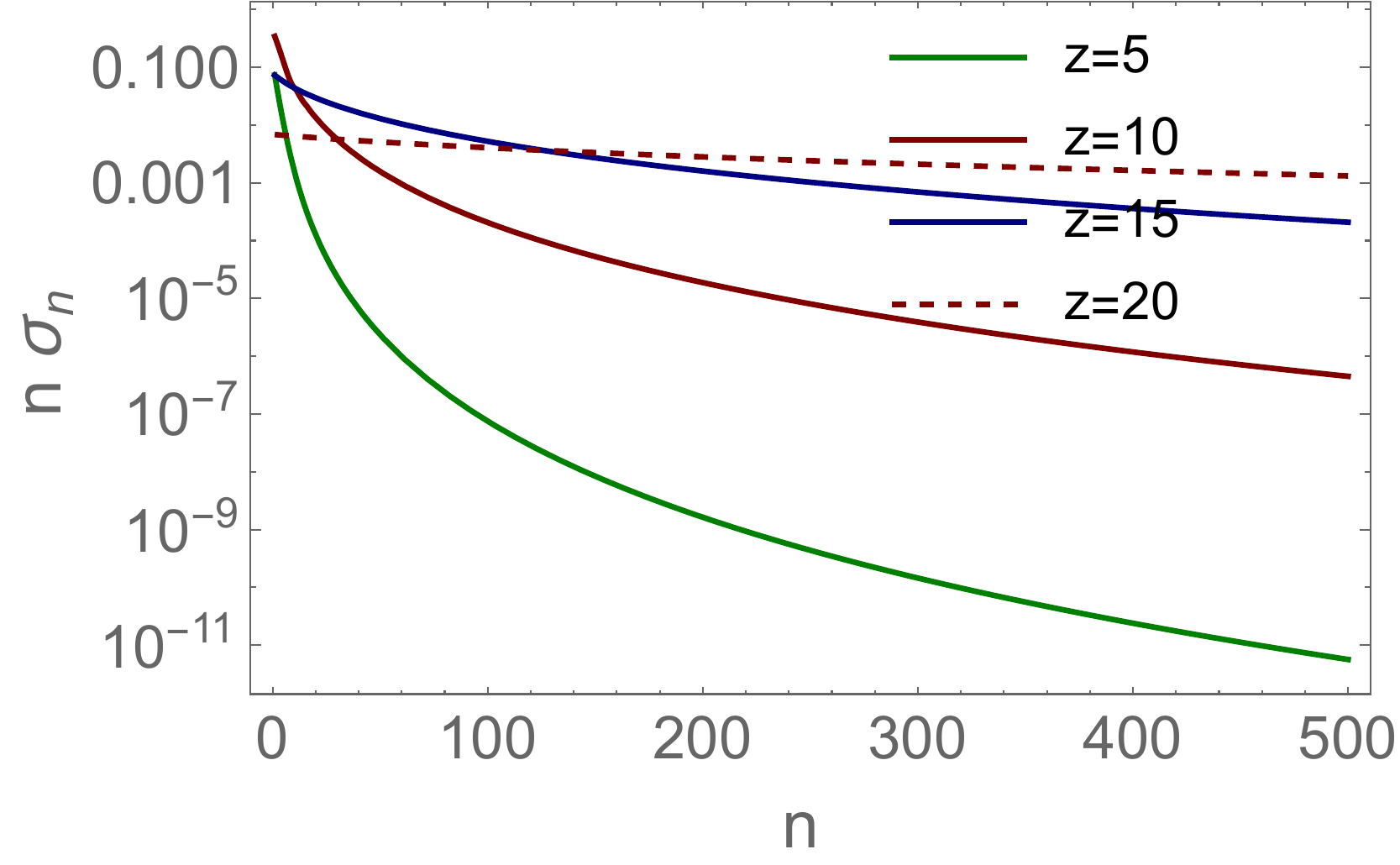}&\includegraphics[width=8.7cm]{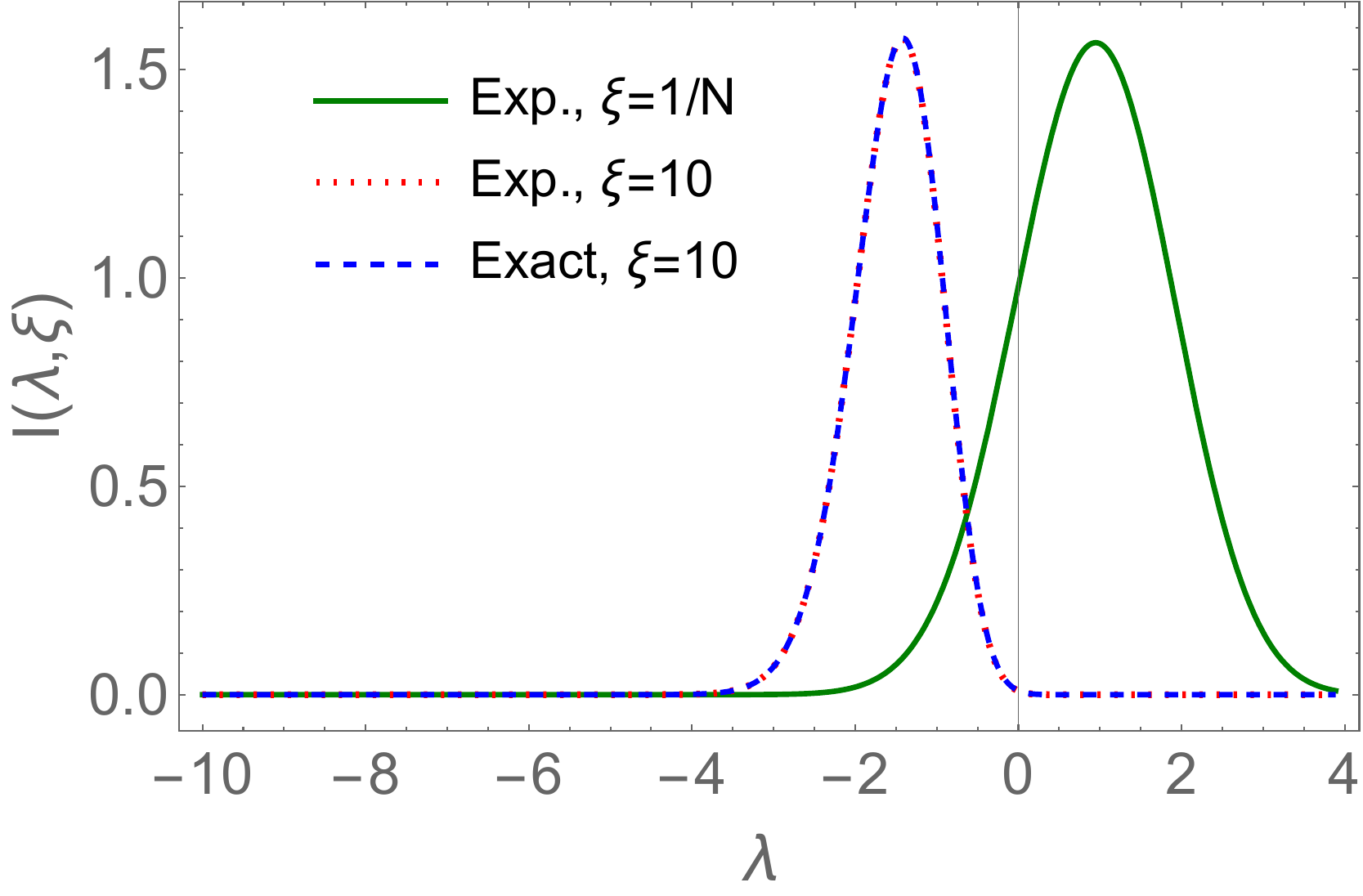} \\
\fig{mdbk}-a & \fig{mdbk}-b\\
\end{tabular}		
		
		\end{center} 
			\caption{\fig{mdbk}-a: Multiplicity distributions for BK scattering process of \eq{MD6}. \fig{mdbk}-b: Integrant of the first term in \eq{MD6} as a function of $\lambda$ at fixed $\xi =\frac{n+1}{N\Lb z_A\Rb}$. For $\xi	=10$\, the integrant is multiplied by factor 200.	For $\xi=1/N$ $N$ is taken to be equal to $N\Lb z_A= 15\Rb$. }
\label{mdbk}
\end{figure}
 %%%%%%%%%%%%%%%%%%%%%%%%%%%%%%%%%%%%%%%%%%%%%%%%%%%%%%%%%%
Plugging $F_k(Y,\xi_{r,R})$ from \eq{MD3} into \eq{AGKK}  and using a new notation: $N\Lb z\Rb =N_0 \exp\Lb \h (z - 4)\Rb$    we obtain:
\begin{subequations}
\bea 
\sigma_n\Lb z\Rb\,&=&\,\sum_{k=n}^\infty\sigma^k_n\Lb z\Rb\,=\,\frac{1}{\sqrt{2\,\pi}}\intl^{\infty}_0\!d t \intl^{\infty}_{-\infty}\!\!\!d \lambda \, e^{ - \h \lambda^2}\sum^\infty_{k=n } (-1)^{k - n} \ \frac{ k!}{(k - n)! \,n!} \Lb \h N\Lb z_A\Rb\Rb^k e^{ - k  \lambda\,-k\,t }   \label{MD4}\\
&=&\frac{1}{\sqrt{2\,\pi}}\intl^{\infty}_0\!d t \intl^{\infty}_{-\infty}\!\!\!d \lambda \, e^{ - \h \lambda^2}\frac{ N^n\Lb z_A\Rb  e^{-n \lambda\,- \,n\,t }}{\Lb 1\,+\, N\Lb z_A\Rb\,e^{ - \lambda\,-\,t }\Rb^{n+1}}
\label{MD5}\\
&= &\frac{1}{\sqrt{2\,\pi}} \frac{1}{n}  \intl^{\infty}_{-\infty}\!\!\!d \lambda \, e^{ - \h \lambda^2}\frac{ N^n\Lb z_A\Rb  e^{-n \lambda\, }}{\Lb 1\,+\, N\Lb z_A\Rb\,e^{ - \lambda }\Rb^{n+1}}
\label{MD6} \\
&= &\frac{1}{\sqrt{2\,\pi}}\Bigg\{\frac{1}{n\,N\Lb z_A\Rb}\!\!\! \intl^{\ln N\Lb z_A\Rb}_{-\infty}\!\!\!\!\!\!d \lambda \, e^{ - \h \lambda^2}\frac{e^{ \lambda}}{ \Lb1 \,+\,\frac{e^\lambda}{N\Lb z_A\Rb}\Rb^{n+1}} +\frac{1}{n} \!\!\! \!\!\intl_{\ln N\Lb z_A\Rb}^{\infty}\!\!\!\!\!\!d \lambda \, e^{ - \h \lambda^2}\,\frac{N^n\Lb z_A\Rb\,e^{  -n\, \lambda}}{\Lb 1\, + \,N\Lb z_A\Rb\,e^{  -\, \lambda}\Rb^{n+1}}\Bigg\}\label{MD6}
\eea
\end{subequations}

The multiplicity distribution of \eq{MD6} is shown in \fig{mdbk}. One can see that $\sigma_n$ falls down at large $n$. This decrease is faster than $1/n$. The  main contribution comes from the first term, which actually we can   interpret as the multiplicity distribution of the first term  ($ 1$)  of the scattering amplitude in \eq{SABK13}. It should be mentioned that factor $1/n$ is not valid in this region as we have discussed in the previous section. Therefore, we believe that we have to take off this factor and discuss \fig{mdbk} as given for $\sigma_n$.

We are going to discuss the first term of \eq{MD6} rewriting it in more convenient form:
\beq\label{MD61}
\underbrace{\frac{1}{N\Lb z_A\Rb}\!\!\!\!\!\! \intl^{\ln N\Lb z_A\Rb}_{-\infty}\!\!\!\!\!\!d \lambda \frac{ e^{ - \h \lambda^2+\lambda}}{ \Lb1 \,+\,\frac{e^\lambda}{N\Lb z_A\Rb}\Rb^{n+1}}}_{Exact}\,=\,
\underbrace{\frac{1}{N\Lb z_A\Rb}\!\!\!\!\!\! \intl^{\ln N\Lb z_A\Rb}_{-\infty}\!\!\!\!\!\!d \lambda \, e^{ - \h \lambda^2+ \lambda}\exp\Lb -(n+1) \frac{e^\lambda}{N\Lb z_A\Rb}\Rb}_{Exp.}\,\,=\underbrace{\frac{1}{N\Lb z_A\Rb}\!\!\!\!\!\! \intl^{\ln N\Lb z_A\Rb}_{-\infty}\!\!\!\!\!\!d \lambda \, e^{ - \h \lambda^2 +\lambda -\xi e^{\lambda}}}_{Exp.}\eeq
In \fig{mdbk}-b we plot the integrants ($I\Lb \lambda,\xi\Rb$) of \eq{MD61}. One can see that (i) $I\Lb \lambda,\xi\Rb$ has  a clear maximum, (ii) the position of the maximum is away of the upper limit of integration ($\ln N\Lb z_A\Rb$) and (iii) the approximate expression (Exp. in \eq{MD61}) works quite well. Based on these features we can state that $\sigma_n$ display the KNO scaling behaviour \cite{KNO}, viz:
\beq\label{MD62}
\sigma_n\,\,=\,\,\frac{1}{N\Lb z_A\Rb} \Psi\Lb \frac{n}{N\Lb z_A\Rb}\Rb
\eeq
  %%%%%%%%%%%%%%%%%%%%%%%%%%%%%%%%%%%%%%%%%%%%%%%%%%%%%%%%%%
 
 \begin{figure}
 	\begin{center}
 	\leavevmode
 		\includegraphics[width=9cm]{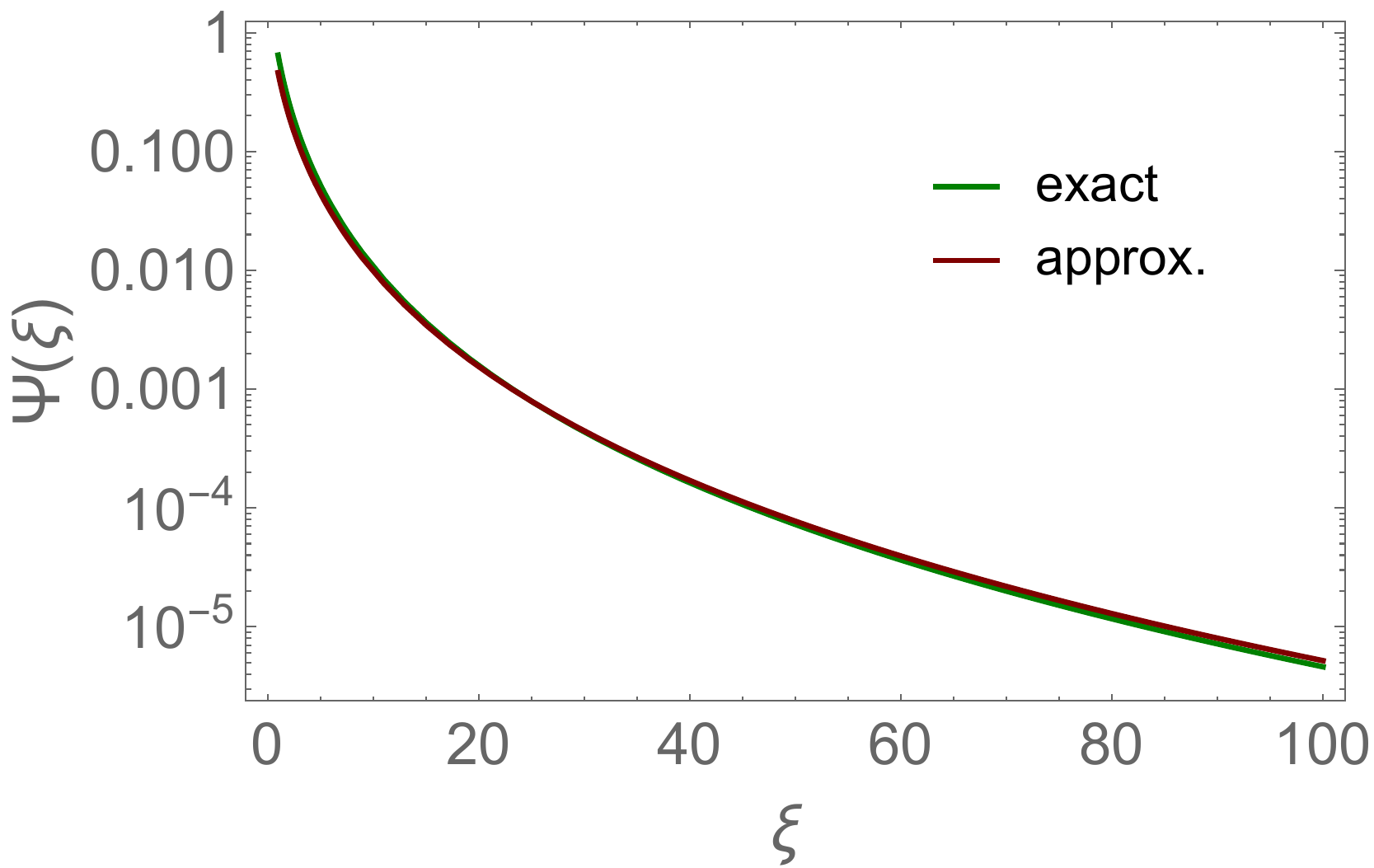}\\
		\end{center} 
			\caption{ The KNO function $\Psi$ versus $\xi =  \frac{n}{N\Lb z_A\Rb}$. Exact is the numerical calculation of \eq{MD61} while Approx. stands for the solution of \eq{MD66}.}		
\label{knobk}
\end{figure}
 %%%%%%%%%%%%%%%%%%%%%%%%%%%%%%%%%%%%%%%%%%%%%%%%%%%%%%%%%%

 We use the steepest decent method to calculate $\Psi$ function with the following equation  for the saddle point in $\lambda$:
\beq\label{MD630}
1 - \lambda_{SP} - \xi e^{\lambda_{SP}}\,=\,0;~~~~\Lb 1 - \lambda_{SP}\Rb\,e^{1 - \lambda_{SP}}\,=\,e \,\xi;
\eeq
The second equation is the Lambert equation for $1 - \lambda_{SP}$ leading to
\beq \label{MD63}
1 - \lambda_{SP}=W\Lb e\,\xi\Rb
\eeq
For Lambert W function we have the series representation:
\beq \label{MD64}
1 -  \lambda_{SP}\,=\,W\Lb e\,\xi\Rb
\,=\,\sum_{n=1}^{\infty} \frac{\Lb - n\Rb^{n-1}}{n!} \Lb e\,\xi\Rb^n
\eeq
 and the asymptotic behaviour  at large values of the argument:
 \beq \label{MD65} 
 W\Lb  e\,\xi\Rb\,\,=\,\,\ln\Lb \frac{e \xi}{\ln(e \xi)} \Rb;~~~
\lambda_{SP}\,=\,- \ln \xi \,+\,\ln\Lb 1 +\ln  \xi\Rb
 \eeq

Using this equation for $\lambda_{SP}$ we obtain:
 \beq \label{MD66}
 \Psi\Lb\xi \equiv  \frac{n}{N\Lb z_A\Rb}\Rb=\sqrt{\frac{\pi}{2}}\exp\Lb - \h \lambda^2_{SP} + \lambda_{SP} - \xi e^{\lambda_{SP}}\Rb
 \eeq
  
 \fig{knobk} shows the KNO function from the exact calculation of the integral in \eq{MD61} and from \eq{MD66}. For large $\xi$ we have $\lambda_{SP} = - \xi$ and
  \beq \label{MD67}
    \Psi\Lb\xi \equiv  \frac{n}{N\Lb z_A\Rb}\Rb\,\,\propto e^{ - \h \xi^2}
    \eeq
  Comparing the multiplicity distribution of \eq{MD62} with the one of Ref.\cite{LEDIDI} we can see that both show the KNO scaling with the same $\xi = n/N(z_A)$. However, the prediction for  $\Psi$ function turns out to be different: \eq{MD66} and \eq{MD67} for this paper and $\Psi = \exp\Lb - \h \xi\Rb$ for Ref. \cite{LEDIDI}.

       ~
   
   ~   
         %%%%%%%%%%%%%%%%%%%%%%%%%%%%%%%%%%%%%%%%%%%%%%%%%%%%  
     \begin{boldmath}
     \section{Dipole-dipole scattering amplitude }
      \end{boldmath}

      %%%%%%%%%%%%%%%%%%%%%%%%%%%%%%%%%%%%%%%%%%%%%%%%%%%%       
   Using \eq{SABK15} we obtain the dipole densities  for fast moving dipole with the  size $r$ and rapidity $Y$ $\rho_n$ in the form:
   \beq \label{RHONFF} 
   \rho_n\Lb Y \{ \vec{r}_i,\vec{b}_i\}\Rb\,\,=\,\, \frac{1}{n} \exp\Lb \beta  \h n^2\,\, -  \,\, 2\alpha\,n\Rb     \prod_{i=1}^n G_{\pom} \Lb z_i\Rb~   \eeq  
   
   with $z_i$ from \eq{zi}.
    
     From \eq{MPSI} we can calculate the dipole-dipole amplitude ($N$) summing the large Pomeron loops. We have for the S-matrix of scattering the dipole with size $r$ and rapidity $Y$ 
      \beq \label{SADD1}
S\Lb z\Rb \,\,=\,\,1\,\,-N\Lb z \Rb\,\,=\,\,\sum^\infty_{n=0} \Lb - 1\Rb^n n! \,C^2_n 
\intl\!\!\! d^2 r_i d^2 r'_i d^2b_i d^2b'_i\prod^n_{i=1} 
\frac{G_{\pom}\Lb z_i\Rb}{N_0}\gamma^{BA}\Lb r_i,r'_i,\delta b\Rb\frac{G_{\pom}\Lb z'_i\Rb}{N_0}\eeq    
      
      Applying the t-channel unitarity to the BFKL Pomeron exchange (see \eq{RHO24}) we  reduce \eq{SADD1} to the following equation:
       \beq \label{SADD2}
S\Lb z\Rb \,\,=\,\,1\,\,-N\Lb z \Rb\,\,=\,\,\sum^\infty_{n=0} \Lb - 1\Rb^n\, n! \,C^2_n\,\Lb \frac{G_{\pom}\Lb z\Rb}{N^2_0}\Rb^n
\eeq
with $z$ from \eq{zz}.

Repeating all steps that we have been discussed in \eq{SABK6}  and  \eq{SABK11} -\eq{SABK12b},  we turn \eq{SADD2} into:
    \beq \label{SADD20}
S\Lb z\Rb \,\,=\,\,1\,\,-N\Lb z \Rb\,\,=\,\,\oint_{C'}\frac{d n}{2\,\pi\,i} \intl^{\infty}_0 dt \,e^{-t}  \, n! \,\,\frac{1}{n^2} e^{ n^2 - 4n}
\Lb \frac{e^{\h \,z}}{N^2_0}\Rb^n\,t^n
\eeq

     Note, that contour $C'$  (see \fig{cont}-a) in this equation does not include the pole at $n=0$ since we consider the S-matrix.  
     
     Taking this integral by the method of steepest descent with the saddle point $n_{SP} = -\frac{1}{4} \,z$, we obtain:
   \beq \label{SADD3}
S\Lb z\Rb \,\,=\,\,\frac{8}{\sqrt{\pi}} \frac{1}{z^2} \exp\Bigg(- \frac{z^2}{16}\,+ z\, - \frac{1}{4}z\Lb \ln\Lb\frac{ z}{4}\Rb - 1\Rb\Bigg)
\eeq
     From this equation one can see that  we reproduce the results for the dipole-dipole scattering amplitude both from the estimates of the rare fluctuation in the CGC \cite{IAMU,IAMU1} and from the BFKL Pomeron calculus in our previous works\cite{LEDIDI,LEDIA,LEAA}. The scattering amplitude that we obtained, turns out to be very close to our approach, based on \eq{I1} (see Refs. \cite{LEDIDI,LEDIA,LEAA}), but it is  more economic and straightforward. The main differences with these papers are  in the way  we sum the asymptotic Pomeron series. 
 As we have discussed, we can calculate $C_n$ only for  large $n$. Therefore, using these $C_n$ the different methods of summation can obtain a different result. Of course, if we know $C_n$ at all values of $n$  any method will lead to the same analytic function. As we have discussed in section IV the approach of this paper
 looks to us more physics motivated. Especially because it passed the check in  the one dimensional model \cite{KLLN} which has exactly calculated scattering amplitude. Actually, the scattering amplitude appears to be very close to the one of Ref.\cite{LEDIDI}, but the multiplicity distribution could lead to a new insight.

      We wish to use the   great advantage of the multi-Pomeron expansion of the scattering amplitude, considering multiplicity distributions of the produced gluons for dipole-dipole scattering processes.

      ~

      ~

            ~

         ~
                                %%%%%%%%%%%%%%%%%%%%%%%%%%%%%%%%%%%%%%%%%%%%%%%%%%%%  
     \begin{boldmath}
     \section{Multiplicity distribution  of produced gluons for dipole-dipole scattering  }
      \end{boldmath}   
   
                                %%%%%%%%%%%%%%%%%%%%%%%%%%%%%%%%%%%%%%%%%%%%%%%%%%%%   
      Using \eq{SADD2} for the dipole-dipole  scattering amplitude with $C_n$ from \eq{SABK9} and rewriting $e^{n^2}$ as follows:
      \beq \label{MDDD1}
      e^{n^2} = 2 \sqrt{\pi}\intl^{\infty}_{-\infty}\!\!\!d \lambda \,e^{ - \frac{\lambda^2}{4} + n \,\lambda}
      \eeq
      we obtain the amplitude in the form:
      \beq \label{MDDD2}
      N\Lb z\Rb\,\,=\,\,2 \sqrt{\pi}\sum_{n=1}^{\infty}\frac{(-1)^{n+1}}{n^2}  \intl^{\infty}_{-\infty}\!\!\! d \lambda \,e^{ - \frac{\lambda^2}{4} + n \lambda}\,n!\,G^n\Lb z'\Rb \,\,=\,\,
2 \sqrt{\pi}\,\sum_{n=1}^{\infty}\frac{(-1)^n}{n^2}  \intl^{\infty}_{-\infty}\!\!\! d \lambda \,e^{ - \frac{\lambda^2}{4} + n \lambda}\,\intl^{\infty}_0\!\! dt \,e^{-t} t^n\,G^n\Lb z\Rb    
\eeq   
In this equation we introduce the integral representation for $\Gamma\Lb n+1\Rb$. 
As we have discussed in the previous section factor $1/n^2$ cannot to be valid at small values of $n$ which contribute to the asymptotic value of $N\Lb z'\Rb = 1$. Indeed, taking this factor off we can sum over $n$ and take the integral with respect to $t$. The answer is
   \beq \label{MDDD3}
      N\Lb z\Rb\,\,=\,\,2 \sqrt{\pi}\, \intl^{\infty}_{-\infty}\!\!\! d \lambda e^{ - \frac{\lambda^2}{4} }\Lb 1 - \frac{e^{1/(e^{\lambda}G(z)} \Gamma \left(0,\frac{1}{(e^{\lambda}G(z))}\right)}{e^{\lambda}G(z)}\Rb
\eeq
  The first term gives 1 f in accord with the $s$ channel unitarity,  while the second leads to the contribution of \eq{SADD3}. Therefore, to reconstruct the multiplicity distribution at large $z$ we need to take off the factor $1/n^2$. However, we have to put it back if we wish to discuss corrections proportional to $\exp\Lb - \frac{z^2}{16}\Rb$.

    Applying the AGK cutting rules of \eq{AGKK},\eq{AGK0}  to this amplitude one can see that $\sigma_n$ is equal to
     
    \bea \label{MDDD4}
\sigma_n\Lb z\Rb & =& 2 \sqrt{\pi}\sum^\infty_{k=n}\Lb - 1\Rb^{k -n }\Lb
 \frac{k!}{(k - n)! \,n!}\Rb\,\intl^{\infty}_{-\infty}\!\!\! d \lambda \,e^{ - \frac{\lambda^2}{4}} \,\intl^{\infty}_0\!\! dt\,\Lb t\, e^{-t+\lambda} \,G\Lb z\Rb \Rb^n \\
 &\,=\,&2 \sqrt{\pi}\,\intl^{\infty}_{-\infty}\!\!\! d \lambda \,e^{ - \frac{\lambda^2}{4}} \intl^{\infty}_0\!\! dt \,e^{-t}\frac{\Lb t\,e^{\lambda} N\Lb z\Rb\Rb^n}{\Lb 1\,+\, t\,e^{\lambda} N\Lb z\Rb\Rb^{n+1}} 
 \,=\,  2 \sqrt{\pi}\,n!\,\intl^{\infty}_{-\infty}\!\!\! d \lambda \,e^{ - \frac{\lambda^2}{4}} 
 \frac{1}{e^{\lambda}\, N\Lb z'\Rb}\,U\left(n+1,1, \frac{1}{e^{\lambda} N\Lb z\Rb}\right)\nn 
 \eea
  where  $U\left(n+1,1, \frac{1}{e^{\lambda} N\Lb z\Rb}\right) $
  is the  Tricomi confluent hypergeometric function (see Ref.\cite{AS}, formula {\bf 13.1.3}).  This function has two interesting limits: 
    
   \bea \label{MDDD5}
\frac{n!}{e^{\lambda}\, N\Lb z\Rb}\,\, U\left(n+1,1,\frac{1}{e^{\lambda}\, N\Lb z\Rb}\right) \,\,\rightarrow\,\,
 \left\{\begin{array}{l}\,\,\,\displaystyle{\frac{1}{e^{\lambda}\, N\Lb z\Rb} \,K_0\Bigg( 2 \sqrt{ \frac{n+1}{e^{\lambda}\, N\Lb z\Rb}}\Bigg)}\,\,\,\,\,\,\,\,\,\,\mbox{for}\,\,\,\,e^{\lambda}\, N\Lb z\Rb\,>\,1;\\ \\
\,\,\displaystyle{\frac{n!}{e^{\lambda}\, N\Lb z\Rb}\Lb n + \frac{1}{e^{\lambda}\, N\Lb z'\Rb}\Rb^{-(n+1)} }\,\,\,\,\,\mbox{for}\,\,\,\,e^{\lambda}\, N\Lb z \Rb\,\,<\,1;\\  \end{array}
\right.
 \eea 
 For these two regions \eq{MDDD4} can be rewritten as follows:
     \bea \label{MDDD6}
\sigma_n\Lb z\Rb & =&   2 \sqrt{\pi}\,\intl^{\infty}_{-\ln N\Lb z\Rb}\!\!\! d \lambda \,e^{ - \frac{\lambda^2}{4}} \displaystyle{\frac{1}{e^{\lambda}\, N\Lb z\Rb} \,K_0\Bigg( 2 \sqrt{ \frac{n+1}{e^{\lambda}\, N\Lb z\Rb}}\Bigg)}\,\,+\,\, 2 \sqrt{\pi}\,\intl^{-\ln N\Lb z'\Rb}_{-\infty}\!\!\! d \lambda \,e^{ - \frac{\lambda^2}{4}}
\displaystyle{\frac{n!}{e^{\lambda}\, N\Lb z\Rb}\Lb n + \frac{1}{e^{\lambda}\, N\Lb z\Rb}\Rb^{-(n+1)} }\eea

 \begin{figure}[ht]
 	\begin{center}
 	\leavevmode
	\begin{tabular}{l l} 
 		\includegraphics[width=8.8cm]{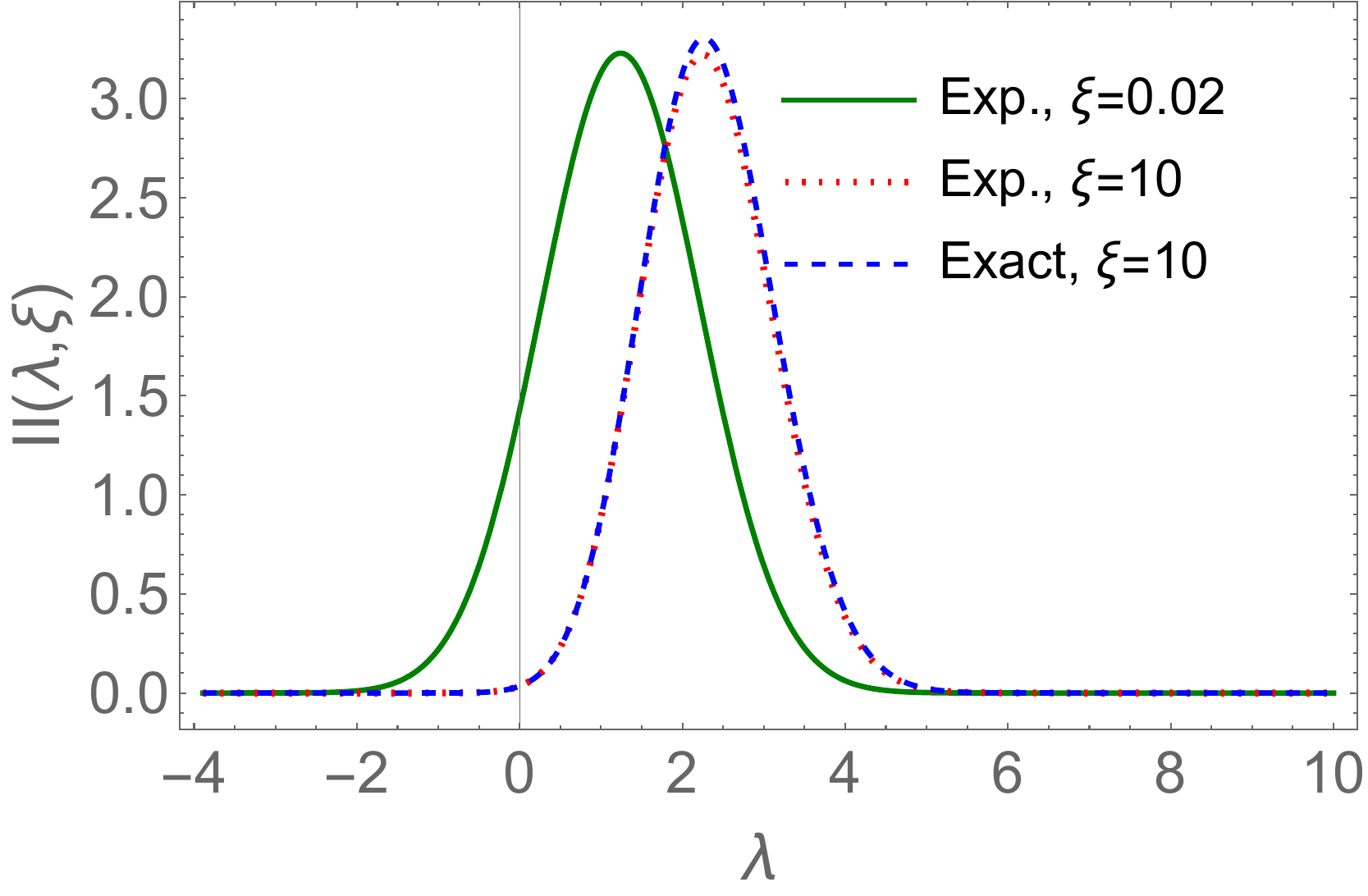}&\includegraphics[width=9.4cm]{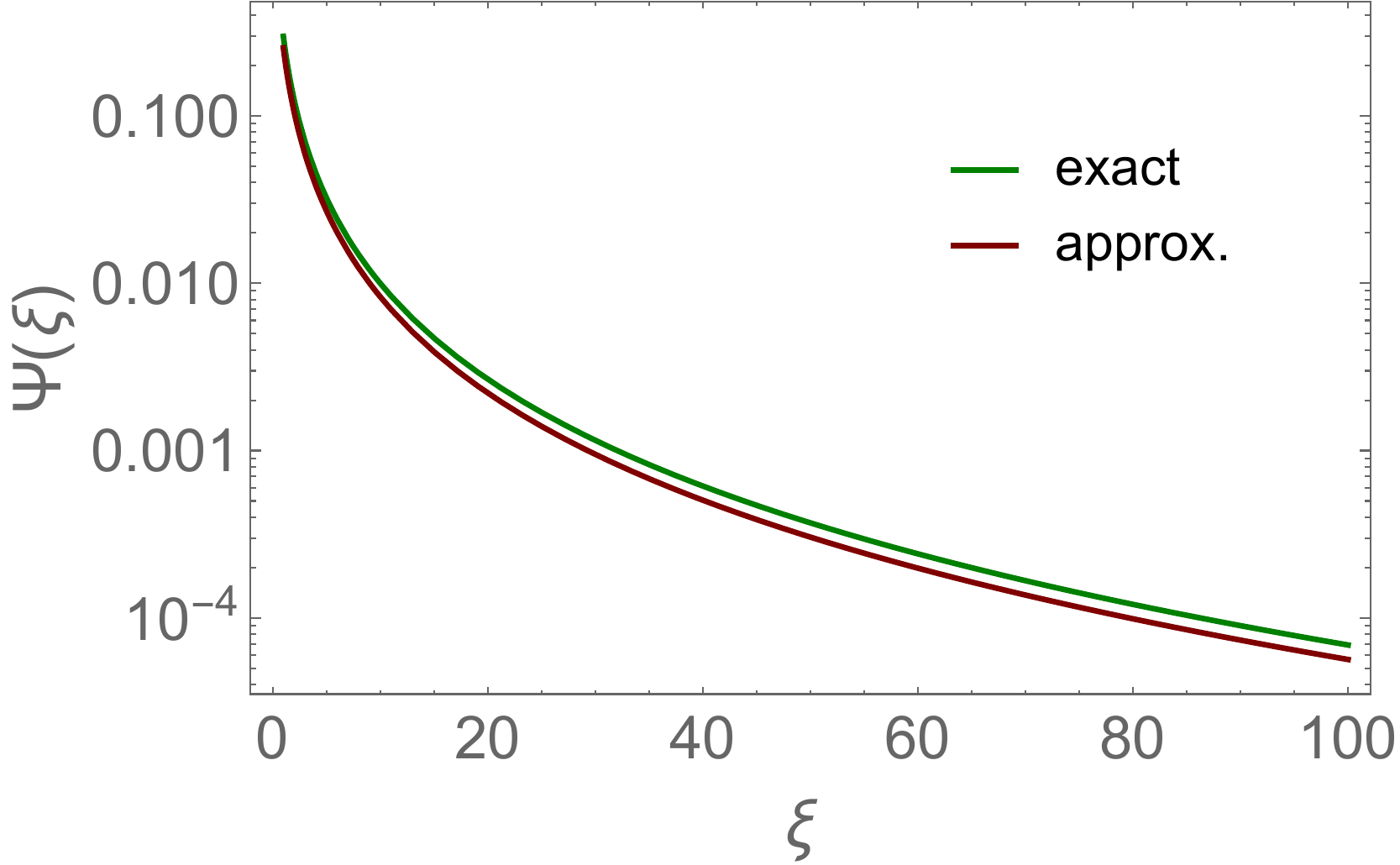} \\
\fig{md}-a & \fig{md}-b\\
\end{tabular}		
		
		\end{center} 
			\caption{Multiplicity distributions for dipole-dipole  scattering process of \eq{MDDD7}. \fig{md}-a: Integrant $II\Lb \lambda, \xi =\frac{n+1}{ N\Lb z'\Rb}\Rb$	of the first term in \eq{MDDD7} as a function of $\lambda$ at fixed $\xi =\frac{n+1}{N\Lb z_A\Rb}$. For $\xi	=10$\, the integrant is multiplied by factor 50.	\fig{md}-b: The KNO function $\Psi$ versus $\xi =  \frac{n}{N\Lb z_A\Rb}$. Exact is the numerical calculation of \eq{MDDD7} while approx. stands for the solution of \eq{MDDD10}			
			 }
\label{md}
\end{figure}

The second term in \eq{MDDD6} gives a small contribution  proportional to $\exp\Lb -  \frac{z'^2}{16}\Rb$. We rewrite the first term in the more convenient form:
\bea \label{MDDD7}
\intl^{\infty}_{-\ln N\Lb z\Rb}\!\!\! \!\!\! \!\!\! d \lambda\,II\Lb \lambda, \xi =\frac{n+1}{ N\Lb z'\Rb}\Rb&=&
 2 \sqrt{\pi}\,\intl^{\infty}_{-\ln N\Lb z'\Rb}\!\!\! \!\!\! \!\!\! d \lambda \,\underbrace{ e^{- \frac{\lambda^2}{4}} \frac{1}{e^{\lambda}\, N\Lb z\Rb} \,K_0\Bigg( 2 \sqrt{\frac{ \xi}{e^{\lambda}}}\Bigg)}_{exact}\nn\\\,
 &\,=\,&\, 2 \sqrt{2\,\pi}\,\intl^{\infty}_{-\ln N\Lb z\Rb}\!\!\! \!\!\! \!\!\! d \lambda \,\underbrace{ e^{- \frac{\lambda^2}{4}} \frac{1}{e^{\lambda}\, N\Lb z\Rb\sqrt{2 \sqrt{\frac{ \xi}{e^{\lambda}}}}} \,\exp\Bigg(- 2 \sqrt{\frac{ \xi}{e^{\lambda}}}\Bigg)}_{approx.}  \eea

\fig{md} shows the $\lambda$ dependence of $II\Lb \lambda, \xi \Rb$ . One can see that 
(i) this dependence has   a clear maximum; (ii) the lower limit of integration does not influence on the value of the integral; and (iii) the approximate expression for $II$ ( see approx. in \eq{MDDD7}) describes the exact integrant quite well. Therefore, we can use the last term in \eq{MDDD7} and take the integral using the method of  steepest descent with the following equation for $\lambda_{SP}$:
\beq \label{MDDD8}
\h\lambda_{SP}+\frac{3}{4} =e^{-\h \lambda_{SP}}\sqrt{\xi};~~\zeta_{SP}\, e^{\zeta_{SP}}= \h e^{\frac{3}{4}}\sqrt{\xi}~~~\mbox{with}~~ \zeta_{SP}\equiv\h \lambda_{SP} + \frac{3}{4}
\eeq
The solution for $\zeta_{SP}$ is the W Lambert function which we have discussed in section V ( see \eq{MD630}-\eq{MD65}). Therefore, the solution is
\beq \label{MDDD9}
 \zeta_{SP}\equiv\h \lambda_{SP} + \frac{3}{4}= W\Lb \h e^{ \frac{3}{4}}\sqrt{\xi}\Rb:~~~~
\lambda_{SP} =\frac{3}{2} + 2W\Lb \h e^{ \frac{3}{4}}\sqrt{\xi}\Rb \eeq

Plugging \eq{MDDD9} into the last equation of \eq{MDDD7} we obtain for the KNO function $\Psi$:
 (see \eq{MD62})
\beq \label{MDDD10}
\Psi\Lb \xi\Rb \,\,=\,\,\displaystyle{\sqrt{2\,\pi}\frac{e^{-\frac{\lambda_{SP} ^2}{4}-2 \sqrt{e^{-\lambda_{SP} } \xi }+\lambda_{SP} }}{\sqrt{2} \sqrt[4]{e^{-\lambda_{SP} } \xi }}}   
\eeq
For large $\xi$ using \eq{MD65} we obtain $ \lambda_{SP} = \xi$ and 
\beq \label{MDDD11}
   \Psi\Lb \xi\Rb \,\,\propto \,\,e^{ - \frac{\xi^2}{4}}
   \eeq
 Comparing \eq{MDDD10} and \eq{MDDD11} with the multiplicity distributions in Ref.\cite{LEDIDI} one can see that they are drastically different. Both  show the KNO scaling   behaviour with $\xi=\frac{n}{N(z)}$  but the shape of function $\Psi$ in Ref.\cite{LEDIDI} is quite different: $\Psi \,\propto \exp\Lb - 2 \sqrt{\xi}\Rb$.  As we have discussed above,  this difference is a consequence of different procedures for summing of the asymptotic series for the scattering amplitude.
   
   ~

   ~
   
                                     %%%%%%%%%%%%%%%%%%%%%%%%%%%%%%%%%%%%%%%%%%%%%%%%%%%%  
     \begin{boldmath}
     \section{Entropy of produced gluons }
      \end{boldmath}

      %%%%%%%%%%%%%%%%%%%%%%%%%%%%%%%%%%%%%%%%%%%%%%%%%%%%  

         We found that for dipole-target  (BK amplitude) and dipole-dipole scattering the multiplicity distributions of produced gluons  have the KNO scaling  behaviour   with different functions $\Psi$ (see \eq{MD66} and \eq{MDDD10} and \fig{md}) but with the same mean multiplicity $\bar{n} =N\Lb z\Rb$\footnote{For BK amplitude $z=z_A$}. This feature is enough to find the entropy at high energies which will be the same 
   in deep inelastic scattering (DIS) and for interactions of two dipoles at high energies.
 Indeed for $\mathcal{P}_n = \frac{1}{\bar{n}}\Psi  \Lb \frac{n}{\bar{n}}\Rb$ the 
      von Neumann entropy is equal 
   \bea \label{EN}
   S_E &= &- \sum_n \ln\Lb\mathcal{P}_n\Rb \, \mathcal{P}_n   =  \ln \bar{n}  \underbrace{\sum_n  \frac{1}{\bar{n}}\Psi  \Lb \frac{n}{\bar{n}}\Rb}_{1}   \,-\,\underbrace{\sum_n\ln\Lb \Psi\Lb \frac{n}{\bar{n}}\Rb  \Rb  \frac{1}{\bar{n}}\Psi  \Lb \frac{n}{\bar{n}}\Rb}_{\rm Const} \nn\\
   &=&\,\,\ln \bar{n} + {\rm Const}\,\,\xrightarrow{z\,\gg\,1} \,\, \ln N\Lb z'\Rb  \,\,=\, \ln\Lb G_{\pom}\Lb z' \Rb\Rb   
   \eea
   Therefore, $S_E \,\,=\,\, \ln N\Lb z_A\Rb =\ln\Lb G_{\pom}\Lb z_A \Rb\Rb  $ in accordance with the result of Ref.\cite{KHLE}. In  \eq{EN} we used the well known fact that the gluon structure function in QCD is equal to the contribution of the exchange of one BFKL Pomeron \cite{MUDI}. Different form of the KNO function $\Psi$ results in a different value of ${\rm Const}$ in \eq{EN}.    
For $\Psi$ of \eq{MD66} this constant is equal to -1.706 while for dipole-dipole scattering its value is -0.217.

   We would like to emphasize that both the value of the entropy and the multiplicity distributions at large $z$ follows directly from QCD    evolution equations and the $t$-channel unitarity constraints of \eq{MPSI}. Bearing this in mind we expect to learn  a lot comparing with other QCD results
(see  Refs.\cite{KHLE,LELAST}), which give a much larger entropy ($\propto\,z^2$) in QCD.
We plan to clarify this discrepancy in our further publications.

It should be noted that the multiplicity distribution are quite different for DIS (BK amplitude) and for dipole-dipole scattering in spite of the same value of the entropy. 
Indeed, we have seen in \eq{MD67} and \eq{MDDD11}
that $\Psi $ for dipole-dipole amplitude is rather equal to $\sqrt{\Psi}$ for BK amplitude at large values of $\xi$. The relation between them at arbitrary $\xi$  is plotted in \fig{knocmp}.

  %%%%%%%%%%%%%%%%%%%%%%%%%%%%%%%%%%%%%%%%%%%%%%%%%%%%%%%%%%
 
 \begin{figure}
 	\begin{center}
 	\leavevmode
 		\includegraphics[width=9cm]{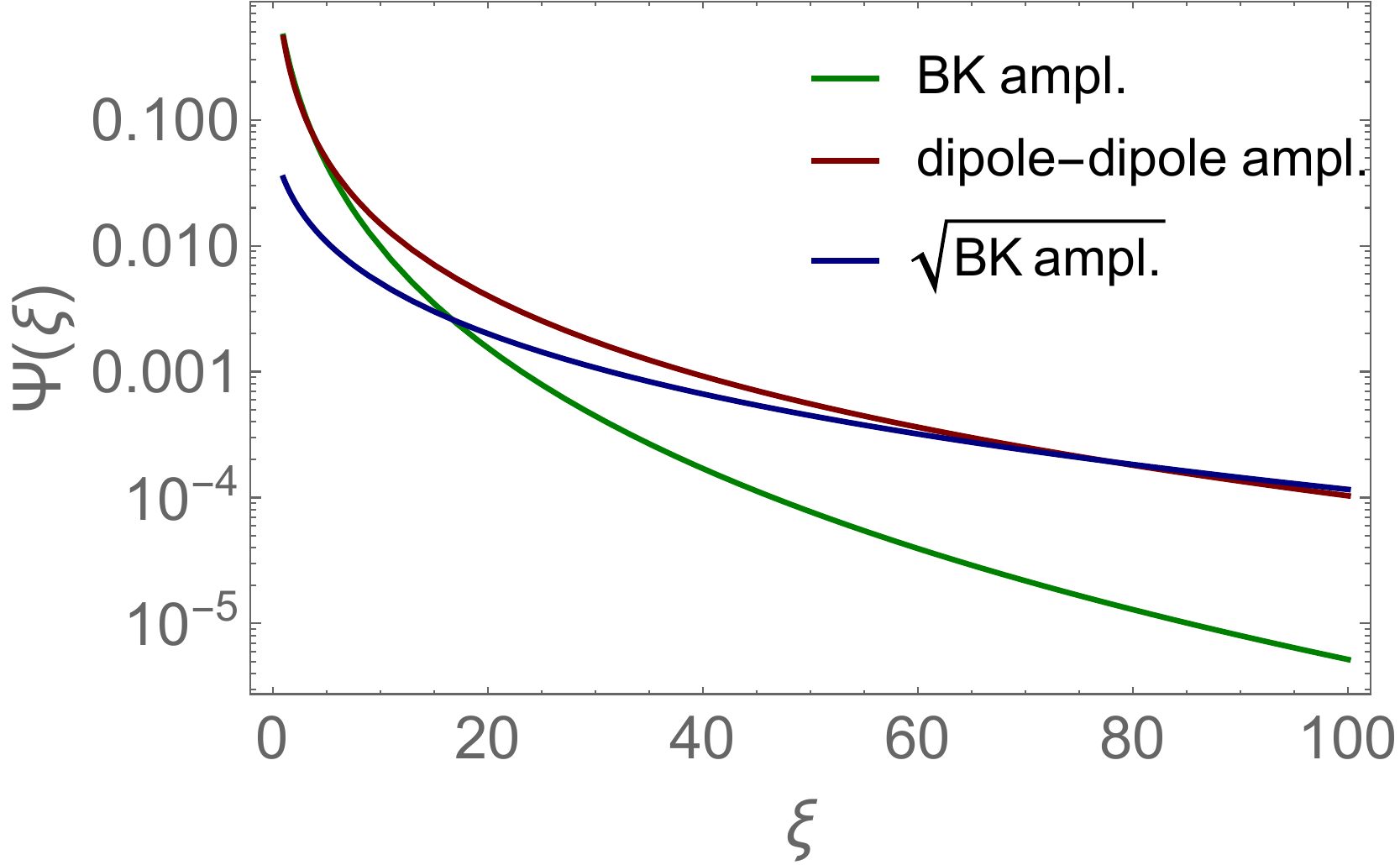}\\
		\end{center} 
			\caption{Comparison of the KNO functions  for BK and dipole-dipole amplitudes. }		
\label{knocmp}
\end{figure}
 %%%%%%%%%%%%%%%%%%%%%%%%%%%%%%%%%%%%%%%%%%%

It is instructive to note that situation turns out to be quite different from the one dimensional models  and for dipole-dipole interaction of Ref.\cite{LEDIDI} in which $\Psi_{dipole-dipole}\Lb \xi\Rb = \Psi_{BK} \Lb 2 \sqrt{\xi}\Rb$ ( see Refs. \cite{MUSA,LEMD1M,LEDIDI}).
      ~

   ~

   ~
   
                                  %%%%%%%%%%%%%%%%%%%%%%%%%%%%%%%%%%%%%%%%%%%%%%%%%%%%  
     \begin{boldmath}
     \section{Conclusions }
      \end{boldmath}

      %%%%%%%%%%%%%%%%%%%%%%%%%%%%%%%%%%%%%%%%%%%%%%%%%%%% 

     In this paper we found the dipole densities (see \eq{RHONFF})  for the leading twist BFKL kernel. These densities stem from the solution of nonlinear Balitsky-Kovchegov evolution equation for the scattering amplitude,
  and $t$-channel unitarity of \eq{MPSI} with the simple expression of \eq{SABK1} for the dipole densities of a nucleus. Recall that \eq{SABK1}  views the nucleus as a bag of dipoles.   Having these densities we use \eq{MPSI} to calculate the contributions of the large Pomeron loops to the scattering amplitude of dipole - dipole interactions at high energies. 
  
  It is shown in the paper  that the QCD equations for dipole densities have a natural solution: the BFKL Pomeron calculus.   We found the scattering amplitudes both for DIS(BK amplitude) and dipole-dipole scattering as the series of the exchanges of BFKL Pomerons and suggested how to sum the resulting asymptotic series.

 All results  for the scattering amplitudes are very close to the approach that we have developed in Refs.\cite{LEDIDI,LEDIA,LEAA}. However, we wish to emphasize  
 that we  actually  have a large difference in the approaches of this paper and of Refs.\cite{LEDIDI,LEDIA,LEAA}. Indeed, in these papers we use two different methods    for summation of  asymptotic  Pomeron series. The approach to this problem in this paper looks more natural  and reliable than in Refs.\cite{LEDIDI,LEDIA,LEAA}. It coincides with the approach that has been developed in one dimensional model \cite{MUSA,KLLN}, in which it  reproduces the exact scattering amplitude. It should be stressed that the model satisfies both $t$ and $s$ unitarity.

  We have to remind that $C_n$  of \eq{RHONFF} have been found for large $n$
  (see \eq{SABK9}  and   \fig{omn}-a). As we have shown (see \fig{omn}-b) we could rely on their value for $z\geq 4$.  The problem to find them at small $n$ has not been solved and has to be postpone  to a future time.
    
   We wish to point out that our amplitude gives the same asymptotic behaviour at large $z$ as  was calculated in Ref.\cite{LETU,IAMU}.
 This result    not only confirms the general idea of Ref.\cite{IAMU} that 'rare' fluctuation contributes to the high energy behaviour of the scattering amplitude but it gives a possibility to calculate its value  in the approach     based on the BFKL Pomeron calculus.

We hope that  our result will stimulate the study of the BFKL Pomeron calculus for finding larger contribution to the scattering matrix. In particular, we have to approach the enhanced diagrams which are crucially dependent on the small Pomeron loops.

    We obtain our S-matrix from the BFKL Pomeron calculus which allow us to study the multiplicity distribution of the produced gluons  
    using AGK cutting rules.   We wish to draw the attention of a reader to the fact that actually we discuss in the paper the multiplicity distributions that correspond to the asymptotic contributions to the scattering amplitudes : $N = 1$.

     It turns out that these secondary gluons have the  KNO  distribution. For both reaction we obtain the same average multiplicity of the produced gluons, which is equal to the gluon structure function deep in the saturation region.    In spite of the fact that the multiplicity distributions have different forms for BK and dipole-dipole amplitude, the same average multiplicity results in the same  entropy which  is equal to $S_E \,=\,\ln(xG(x,Q^2))$ where $xG$ is the gluon structure function. Therefore it confirms the result of Ref.\cite{KHLE}. It is instructive to note that the KNO function for dipole-dipole amplitude turns out to $\Psi_{dipole-dipole}\Lb \xi\Rb = \sqrt{\Psi_{BK}\Lb \xi\Rb}$ at large $\xi$ in remarkable difference from one dimensional models and Ref.\cite{LEDIDI}  where  $\Psi_{dipole-dipole}\Lb \xi\Rb = \Psi_{BK} \Lb \sqrt{\xi}\Rb$ ( see Refs. \cite{MUSA,LEMD1M}).

    We hope that this paper will contribute to the further study of the Pomeron calculus in QCD.

    ~

    ~

       {\bf Acknowledgements} 
     
   We thank our colleagues at Tel Aviv university  for
 discussions. Special thanks go A. Kovner and M. Lublinsky for stimulating and encouraging discussions on the subject of this paper. 
  This research was supported  by 
BSF grant 2022132.

~

~

~

\end{document}